\documentclass[12pt,dvips]{article}
\usepackage{rotating}
\usepackage{epsfig}
\usepackage{color}
\usepackage{cite}

\usepackage{hhline}
\usepackage{multirow}

\begin{document}

\def\thefootnote{\fnsymbol{footnote}}

%%%%%%%%%%%%%%%%%%%%%% F I G U R E %%%%%%%%%%%%%%%%%%%%%%%%%%%%%%%%%%%
\begin{figure*}[t]
\begin{flushleft}
\resizebox{2cm}{!}{\includegraphics{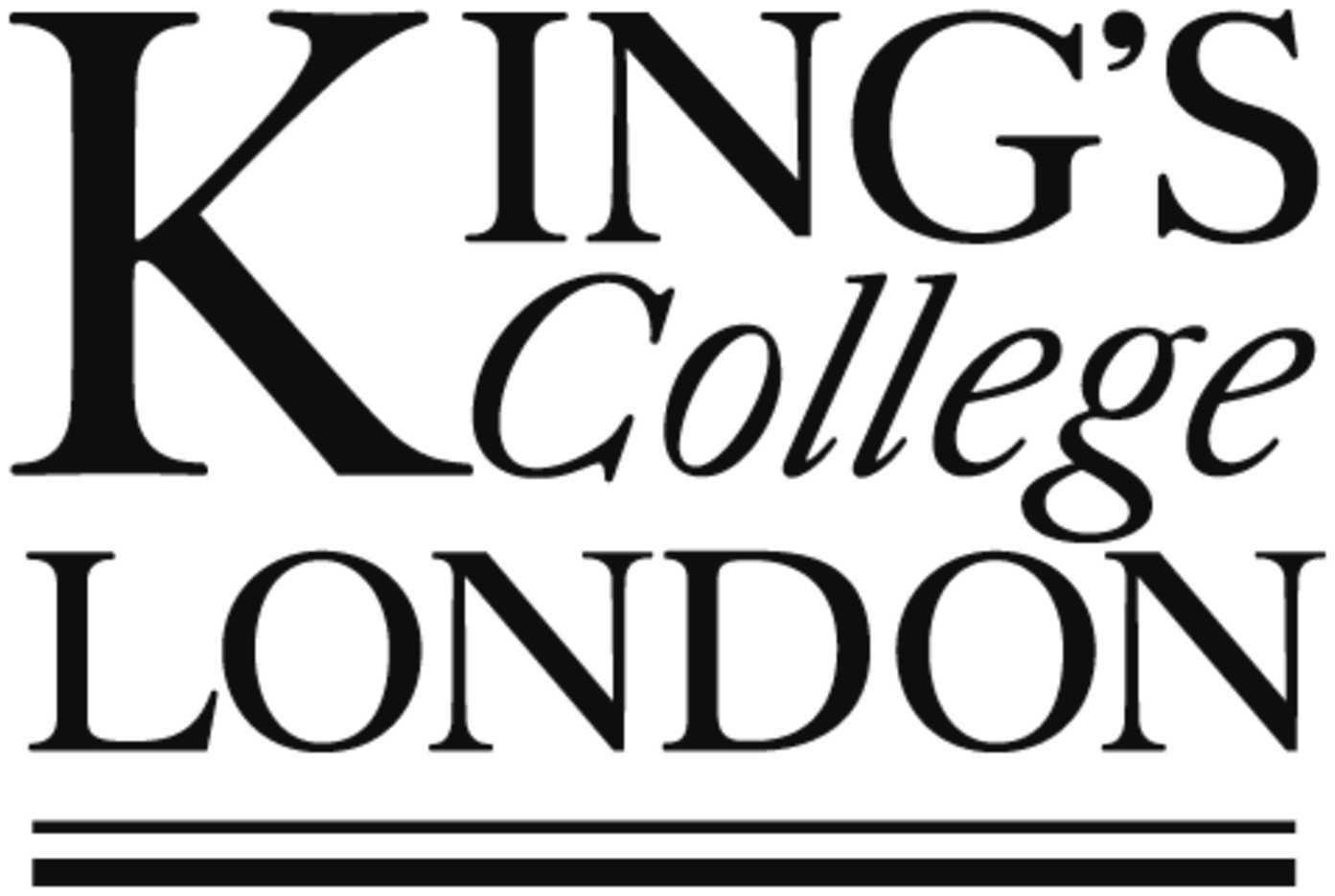}}
\end{flushleft}
\end{figure*}
${}$~\vspace{-4.8cm}
\begin{flushright}
{\tt CERN-PH-TH/2011-008}, {\tt KCL-PH-TH/2011-02}, {\tt MAN/HEP/2011/01}\\
{~}\\
%{\tt arXiv:1006.NNNN} \\
%\vspace{1cm}
%\today
January 2011
%Last modified by JE on \today
\end{flushright}
%%%%%%%%%%%%%%%%%%%%%% F I G U R E %%%%%%%%%%%%%%%%%%%%%%%%%%%%%%%%%%%

\vspace{1cm}
\begin{center}
{\bf {\LARGE Maximal Electric Dipole Moments of Nuclei\\[3mm] with
    Enhanced Schiff Moments}}
\end{center}

\medskip

\begin{center}{\large
John~Ellis$^{a,b}$,
Jae~Sik~Lee$^c$ and
Apostolos~Pilaftsis$^d$}
\end{center}

\begin{center}
{\em $^a$Theory Division, CERN, CH-1211 Geneva 23,
  Switzerland}\\[0.2cm]
{\em $^b$Theoretical Particle Physics and Cosmology Group, Department of
  Physics, King's~College~London, London WC2R 2LS, United Kingdom}\\[0.2cm] 
{\em $^c$Physics Division, National Center for Theoretical Sciences, 
Hsinchu, Taiwan 300}\\[0.2cm]
{\em $^d$School of Physics and Astronomy, University of Manchester,}\\
{\em Manchester M13 9PL, United Kingdom}
\end{center}

\bigskip\bigskip

\centerline{\bf ABSTRACT}

\noindent  
The  electric   dipole  moments  (EDMs)  of  heavy   nuclei,  such  as
${}^{199}{\rm Hg}$, ${}^{225}{\rm Ra}$  and ${}^{211}{\rm Rn}$, can be
enhanced  by the  Schiff moments  induced  by the  presence of  nearby
parity-doublet states.  Working within  the framework of the maximally
CP-violating and  minimally flavour-violating (MCPMFV)  version of the
MSSM, we discuss the maximal values that such EDMs might attain, given
the  existing experimental  constraints on  the Thallium,  neutron and
Mercury EDMs.  The maximal EDM values of the heavy nuclei are obtained
with the help of a differential-geometrical approach proposed recently
that  enables  the  maxima  of  new  CP-violating  observables  to  be
calculated  exactly  in the  linear  approximation.   In  the case  of
$^{225}$Ra, we find  that its EDM may  be as large as 6  to $50 \times
10^{-27}$~$e\cdot {\rm cm}$.


\newpage

\section{Introduction}
\label{sec:intro}

Electric dipole moments (EDMs)  are among the most promising potential
signatures for  CP-violating physics  beyond the Standard  Model (SM),
and one of the most promising options for extending the SM is provided
by supersymmetry (SUSY)~\cite{nath}. The minimal SUSY extension of the
SM (the MSSM) already contains many possible CP-violating phases, even
in  its  minimally  flavour-violating  (MFV) version.   The  maximally
CP-violating  MFV version  of  the  MSSM, the  MCPMFV  model~\cite{Ellis:2007kb,RefMFV}, has  six
CP-violating  phases, to  which  may  be added  the  QCD vacuum  phase
$\theta_{\rm  QCD}$.  These  phases  are tightly  constrained  by  the
present  experimental upper  limits  on the  EDMs  of $^{205}$Tl,  the
neutron and $^{199}$Hg.  Nevertheless, one could in principle envisage
(accidental) cancellations~\cite{IN}  between the contributions  to the measured
EDMs  of  the  six  (seven)  phases of  the  MCPMFV  model  (including
$\theta_{\rm QCD}$),  which might leave open the  possibility of large
net contributions to other EDMs.  However, for any fixed values of the
CP-conserving   MCPMFV   parameters,  the   compact   ranges  of   the
CP-violating parameters imply that the value of any other CP-violating
observable, e.g.,  an EDM, is  necessarily bounded. The  question then
arises whether the prospective  sensitivity of any proposed experiment
reaches below  the maximum value  attainable in any  given theoretical
framework, such as  the MCPMFV model (with or  without the possibility
that  $\theta_{\rm QCD}  \ne 0$).   Clearly, any  experiment  that has
insufficient  sensitivity to  search below  the maximum  value  is not
interesting  for  testing the  MCPMFV  model,  whereas any  experiment
capable  of  reaching  below  the  maximum value  may  either  make  a
measurement or exclude part of the MCPMFV parameter space.

In  a  recent paper~\cite{ELPouter}  we  proposed  a novel  analytical
technique,  based  on  a  differential-geometrical  construction  (see
also~\cite{ELPmath}), for  finding the maximal  values of CP-violating
observables subject to the existing EDM constraints, which is exact in
the linear  approximation.  We applied this technique  to find maximal
values  of  the  EDMs  of  the Deuteron  and  muon,  the  CP-violating
asymmetry in  $b \to  s\gamma$ decay, $A_{CP}$,  and the  $B_s$ mixing
phase~\cite{ELPouter}. We found that,  whereas the EDM of the Deuteron
in the MCPMFV model (allowing also for $\theta_{\rm QCD} \ne 0$) might
be  one  (two)  orders   of  magnitude  larger  than  the  prospective
experimental sensitivity,  and $A_{CP}$ might also  be detectable, the
EDM of the muon and the  contribution to the $B_s$ mixing phase in the
MCPMFV model are  likely to be too small to be  observable in the near
future.

In  this   paper  we  extend   the  applications  of   our  analytical
method~\cite{ELPouter,ELPmath} to calculate  the maximal values in the
MCPMFV of  the EDMs  of some  nuclei that are  enhanced by  the Schiff
moment    contributions   associated   with    nearby   parity-doublet
states~\cite{Schiff}. An experimental campaign is now being considered
for   HIE-ISOLDE    that   could    search   for   EDMs    of   radium
isotopes~\cite{Willmann},  accompanied  by  measurements  of  octupole
collectivity in radium isotopes~\cite{Butler}  that would be needed to
interpret  EDM  measurements   in  terms  of  time-reversal  violating
interactions.

We find that values of the $^{225}$Ra EDM that are considerably larger
than $10^{-27}$~$e  \cdot {\rm cm}$ (well within the  estimated sensitivity
of  the proposed  HIE-ISOLDE experiment)  are possible  in  the MCPMFV
model. How much larger depends quite sensitively on the implementation
of the experimental constraint on  the EDM of $^{199}$Hg. As discussed
below, several theoretical calculations  of the EDM are available, and
most  of  them  give  very  similar  allowed ranges  for  the  EDM  of
$^{225}$Ra, in the range 6  to $10 \times 10^{-27}$~$e\cdot {\rm cm}$.
However, one theoretical calculation of the $^{199}$Hg EDM yields much
weaker constraints  on the CP-violating parameters of  the MCPMFV, and
hence yields a much larger maximal value of the $^{225}$Ra EDM, namely
$50 \times 10^{-27}$~$e \cdot {\rm cm}$.

\section{Schiff Moments for Selected Nuclei}
\label{sec:schiff}

A  CP-violating  atomic EDM  may  arise  from  intrinsic EDMs  of  the
constituent  nucleons and  atomic  electrons, CP-odd  electron-nucleon
interactions,  and  the CP-odd  nuclear  moment  known  as the  Schiff
moment.  The Schiff  moments of several atoms have  been calculated in
the literature, and expressed in  terms of sums of products of CP-even
and CP-odd pion-nucleon-nucleon ($\pi  NN$) couplings.  We recall that
the    CP-even   strong    $\pi    N   N$    interaction   is    given
by~\cite{Khriplovich:1999qr}
\begin{equation}
{\cal L}_{\pi NN}^{\rm strong}\ =\
g_{\pi NN}\left[
\sqrt{2} \left(\bar{p} i\gamma_5 n \pi^+\: +\: \bar{n} i\gamma_5 p \pi^-\right)
\: +\: \left(\bar{p}i\gamma_5 p-\bar{n}i\gamma_5 n\right)\pi^0\right] ,
\end{equation}
where $g_{\pi NN}=13.45$.  On the  other hand, the CP-odd (T-odd) $\pi
NN$                interactions                are               given
by~\cite{Khriplovich:1999qr,Pospelov:2005pr}
\begin{eqnarray}
{\cal L}_{\pi NN}^{\rm T\!\!\!\!/}
&=& \bar{g}^{(0)}_{\pi NN} \overline{N}\tau^a N \pi^a \ + \
\bar{g}^{(1)}_{\pi NN} \overline{N} N \pi^0 \ + \
\bar{g}^{(2)}_{\pi NN} \left(
\overline{N}\tau^a N \pi^a - 3\overline{N} \tau^3 N \pi^0\right)
\nonumber \\
&=& \ \bar{g}^{(0)}_{\pi NN} \left[
\sqrt{2} \left(\bar{p} n \pi^+ + \bar{n}p \pi^-\right) +
\left(\bar{p}p-\bar{n}n\right)\pi^0\right]
\nonumber \\
&&\hspace{-0.3cm}
+\ \bar{g}^{(1)}_{\pi NN} \left( \bar{p}p+\bar{n}n\right)\pi^0
\nonumber \\
&&\hspace{-0.3cm}
+\  \bar{g}^{(2)}_{\pi NN} \left[
\sqrt{2} \left(\bar{p} n \pi^+ + \bar{n}p \pi^-\right) -2
\left(\bar{p}p-\bar{n}n\right)\pi^0\right]
\label{eq:cpodd_pinn}
\end{eqnarray}
in  terms   of  the  isoscalar   $\bar{g}^{(0)}_{\pi  NN}$,  isovector
$\bar{g}^{(1)}_{\pi  NN}$,   and  isotensor  $\bar{g}^{(2)}_{\pi  NN}$
T-violating pion-nucleon couplings.

The  Schiff  moment  is  linear  in  the  CP-odd  $\pi  NN$  couplings
$\bar{g}^{(i)}_{\pi NN}$, and may be written as~\cite{Ban:2010ea}
\begin{equation}
S\ =\ (a_0+b) \, g_{\pi NN} \bar{g}^{(0)}_{\pi NN} \: + \:
 a_1 \, g_{\pi NN} \bar{g}^{(1)}_{\pi NN} \: + \:
 (a_2-b) \, g_{\pi NN} \bar{g}^{(2)}_{\pi NN} .
\label{eq:Schiff}
\end{equation}
Here  the coefficients  $a_i$  specify the  dependence  of the  Schiff
moment on  the CP-odd interactions  and the coefficient  $b$ specifies
its dependence on the  nucleon dipole moments.  The coefficients $a_i$
with $i=0,1,2$  and $b$ depend  on the type  of atom of  interest, and
some theoretical estimates  for $^{199}$Hg, $^{225}$Ra and $^{211}$Rn,
in    units   of    $e   \cdot    {\rm   fm}$    are    collected   in
Table~\ref{tab:Schiff}.

%\begin{table}[htb!]
\begin{table}[t!]
\caption{\it The coefficients $a_i, i = 0, 1, 2$ and $b$ of the Schiff
  moments  of  $^{199}$Hg, $^{225}$Ra,  and  $^{211}$Rn, expressed  in
  units  of $e\cdot{\rm  fm}^3$.   The  labels HB  and  HFB stand  for
  calculations   in  the   Hartree-Fock   and  Hartree-Fock-Bogoliubov
  approximations,   respectively:   see   Ref.~\cite{Ban:2010ea}   for
  details.   We have  changed  the signs  of  the coefficients  $a_0$,
  $a_1$,     and    $b$     to    follow     the     conventions    of
  Ref.~\cite{Pospelov:2005pr}. }
\begin{center}
\begin{tabular}{ l||rlrrrr }
\hline\hline
Atom & Ref. & Interaction & $-a_0~~$ & $-a_1~~$ & $a_2~~$ & $-b~~$  \\
\hline\hline 
%-- 199Hg
\multirow{8}{*}{$^{199}$Hg} &
\cite{Dmitriev:2003kb}& --- & $0.0004$ & $0.055$ & $0.009$ & ---  \\
\hhline{~------}
& \cite{deJesus:2005nb} & SkO$^\prime$ & $0.010$ & $0.074$ & $0.018$ & ---  \\
& & (average) & $0.007$ & $0.071$ & $0.018$ & ---  \\
\hhline{~------}
& \cite{Ban:2010ea} & SLy4 (HF) & $0.013$ & $-0.006$ & $0.022$ & $0.003$ \\
& & SIII (HF) & $0.012$ & $0.005$ & $0.016$ & $0.004$ \\
& & SV (HF) & $0.009$ & $-0.0001$ & $0.016$ & $0.002$ \\
& & SLy4 (HFB) & $0.013$ & $-0.006$ & $0.024$ & $0.007$ \\
& & SkM$^*$ (HFB) & $0.041$ & $-0.027$ & $0.069$ & $0.013$ \\
\hline\hline
%-- 225Ra
\multirow{2}{*}{$^{225}$Ra} &
\cite{Engel:2003rz}& SkO$^\prime$ (zero range)& $-5.1$ & $10.4$ & $-10.1$ & --- \\
\hhline{~------}
& \cite{Dobaczewski:2005hz} & SkO$^\prime$ & $-1.5$ & $6.0$ & $-4.0$ & --- \\
\hline\hline
%-- 211 Rn
\multirow{4}{*}{$^{211}$Rn} &
\cite{Dmitriev:2004fk}
&  & $0.019$ & $-0.061$ & $0.053$ & --- \\
\hhline{~------}
& \cite{Ban:2010ea}
& SLy4  & $0.042$ & $-0.018$ & $0.071$ & $0.016$ \\
& & SkM$^*$  & $0.042$ & $-0.028$ & $0.078$ & $0.015$ \\
& & SIII  & $0.034$ & $-0.0004$ & $0.064$ & $0.015$ \\
\hline\hline
\end{tabular}
\end{center}
\label{tab:Schiff}
\end{table}

In    the    case   of    the    $^{199}$Hg    Schiff   moment,    the
results~\cite{Dmitriev:2003kb}                                  adopted
in~\cite{Pospelov:2005pr,Ellis:2008zy,cpsuperh}    are   significantly
different       from      the      more       recent      calculations
of~\cite{deJesus:2005nb,Ban:2010ea}.   In particular,  the coefficient
$a_0$ may be enhanced by a factor of $\sim 30$, whilst the coefficient
$a_1$ may be reduced by a factor  of $\sim 10$ and could even take the
opposite sign.  After considering  the different numbers from the most
recent  calculations~\cite{Ban:2010ea}  in Table~\ref{tab:Schiff},  we
have used  the SIII (HF)  calculation for our  numerical illustration.
As motivation, we note that the  first SLy4 and the SV calculations do
not yield  the right  quantum numbers for  the ground state,  that the
second SLy4 calculation was found  to be unable to project HFB states,
and that the SkM* calculation gives a spherical minimum.

The  CP-odd couplings  $\bar{g}^{(i)}_{\pi  NN}$ may  be generated  by
chromoelectric   dipole   moments  (CEDMs)   of   the  quarks   and/or
dimension-six          four-fermion          interactions.          In
Ref.~\cite{Pospelov:2001ys},  one may  find full  expressions  for the
contributions  to the  CP-odd isoscalar  $\bar{g}^{(0)}_{\pi  NN}$ and
isovector  $\bar{g}^{(1)}_{\pi NN}$  couplings from  the CEDMs  of the
light   quarks~\footnote{We  adopt   the  conventions   and  notations
  of~\cite{Ellis:2008zy,cpsuperh}.}:
\begin{eqnarray}
\bar{g}^{(0)}_{\pi NN} & = &
(-0.5 ~{\rm to}~ 1.5) \times 10^{-12} \,
\frac{(d^C_u+d^C_d)/g_s}{10^{-26}\,{\rm cm}}\,
\frac{\left|\langle\bar{q}q\rangle\right|}{(225\,{\rm MeV})^3}\,, \nonumber \\
\bar{g}^{(1)}_{\pi NN} & = &
2^{+4}_{-1} \times 10^{-12} \,
\frac{(d^C_u-d^C_d)/g_s}{10^{-26}\,{\rm cm}}\,
\frac{\left|\langle\bar{q}q\rangle\right|}{(225\,{\rm MeV})^3}\,. 
\end{eqnarray}
We note that  a `best value' is not  available for $\bar{g}^{(0)}_{\pi
  NN}$.    For  definiteness,   we   have  taken   $\bar{g}^{(0)}_{\pi
  NN}/\bar{g}^{(1)}_{\pi  NN}=0.2\,  (d^C_u+d^C_d)/(d^C_u-d^C_d)$,  as
follows from assuming that  the couplings $\bar{g}^{(0)}_{\pi NN}$ and
$\bar{g}^{(1)}_{\pi NN}$  are proportional  to the matrix  elements of
$\bar{u}u-\bar{d}d$  and  $\bar{u}u+\bar{d}d$  in the  nucleon  state,
respectively.   It is  known that  the matrix  element with  the minus
($-$) sign  is smaller than the  one with the  plus ($+$) sign by  a a
factor $\sim$ 5 to 10~\cite{private:PospelovRitz}.

We have also included
the contribution of the dimension-six four-fermion interactions to the
isovector coupling which can be enhanced for a large value of
$\tan\beta$~\cite{Lebedev:2002ne,Demir:2003js}:
\begin{eqnarray}
\bar{g}^{(1)}_{\pi NN} & = &
\frac{\langle\bar{q}q\rangle}{2f_\pi}\,
\langle N| C_{dd}\,\bar{d}d+
C_{sd}\,\bar{s}s+ C_{bd}\,\bar{b}b |N \rangle  \nonumber \\
& =&
-8\times 10^{-3} {\rm GeV}^3\,
\left[\frac{0.5C_{dd}}{m_d}+3.3\kappa\frac{C_{sd}}{m_s}
+(1-0.25\kappa)\frac{C_{bd}}{m_b} \right]\, ,
\label{4fermion}
\end{eqnarray}
where the  couplings appearing in (\ref{4fermion}) are  defined via
the interaction Lagrangian
\begin{equation}
  \label{L4f}
{\cal     L}_{\rm      4f}\ =\  \sum_{f,f'}     C_{ff'}     (\bar{f}f)
(\bar{f'}i\gamma_5f')\; ,
\end{equation}
and $\kappa\equiv\langle N  | m_s \bar{s} s |  N \rangle/220~{\rm MeV}
\simeq   0.50\pm0.25$.    The   contribution   of   the   four-fermion
interactions  to the  isoscalar coupling  $\bar{g}^{(0)}_{\pi  NN}$ is
ignored  and the  isotensor coupling  $\bar{g}^{(2)}_{\pi  NN}$, which
changes the  isospin by two  units, is neglected  in this work  (as in
Ref.~\cite{Pospelov:2001ys}), since  it can  be generated only  at the
expense of an additional $m_u-m_d$ suppression.

In  the  following, we  concentrate  on  the  EDMs of  $^{199}$Hg  and
$^{225}$Ra atoms,  because there  is already a  stringent experimental
upper limit  on the  former, and  there is a  proposal to  measure the
latter  at  HIE-ISOLDE~\cite{Willmann}.  Moreover,  we  note that  the
estimates  of the  contributions to  the Schiff  moment  of $^{211}$Rn
shown in  Table~\ref{tab:Schiff} are considerably  less favourable. 

The      experimental      constraint      on     the      EDM      of
$^{199}$Hg~\cite{Romalis:2000mg,Griffith:2009zz}, 
\begin{equation}
\left|d_{\rm Hg}^{\rm EXP}\right|\ <\ 3.1\times 10^{-29}\, e\cdot {\rm
  cm}\ (95\,\%\,{\rm C.L.}), 
\end{equation}
is among  the most  stringent constraints on  CP violation  imposed by
atomic  EDM  experiments.   There  are  several  calculations  of  the
$^{199}$Hg EDM in terms of  the Schiff moment in the literature, which
can be cast in the general form
\begin{eqnarray}
d_{\rm Hg}[S] &= &\,
 10^{-17}\ {\cal C}_{\rm Hg}^S\ e\cdot {\rm cm} \times
\left(\frac{S}{e\cdot {\rm fm}^3}\right)\; ,
\end{eqnarray}
with    the   coefficient   taking    the   values    ${\cal   C}_{\rm
  Hg}^S=-4$~\cite{Flambaum:1985ty,Flambaum:1985gv},
$-2.8$~\cite{Dzuba:2002kg}  and $+5.07$~\cite{Latha:2009nq}.   We note
that we have changed the sign  of the most recent calculation based on
relativistic  coupled-cluster  theory~\cite{Latha:2009nq},  so  as  to
match        the       conventions        in        the       previous
calculations~\cite{Flambaum:1985ty,Flambaum:1985gv,Dzuba:2002kg} based
on  coupled perturbed  Hartree-Fock calculations,  and that  it  has a
different sign.   For our numerical calcluatons we  take the estimates
of the coefficient ${\cal C}_{\rm Hg}^S$ that have negative signs.  To
be  specific, we  have used  the  following four  calculations of  the
Mercury EDM:
\begin{itemize}

\item  From Ref.~\cite{Pospelov:2005pr},
\begin{equation}
\label{eq:dhg1}
d^{\rm \,I}_{\rm Hg}[S]\ \simeq\
1.8 \times 10^{-3}\, e\,\bar{g}^{(1)}_{\pi NN}\,/{\rm GeV}\,,
\end{equation}

\item  Taking ${\cal C}^S_{\rm Hg}=-2.8$~\cite{Dzuba:2002kg}
and the coefficients from Ref.~\cite{Dmitriev:2003kb},
\begin{equation}
\label{eq:dhg2}
d^{\rm \,II}_{\rm Hg}[S]\ \simeq\
7.6 \times 10^{-6}\, e\,\bar{g}^{(0)}_{\pi NN}\,/{\rm GeV}+
1.0 \times 10^{-3}\, e\,\bar{g}^{(1)}_{\pi NN}\,/{\rm GeV}\,,
\end{equation}

\item  Taking ${\cal C}^S_{\rm Hg}=-2.8$~\cite{Dzuba:2002kg}
and the average coeffcients from Ref.~\cite{deJesus:2005nb},
\begin{equation}
\label{eq:dhg3}
d^{\rm \,III}_{\rm Hg}[S]\ \simeq\
1.3 \times 10^{-4}\, e\,\bar{g}^{(0)}_{\pi NN}\,/{\rm GeV}+
1.4 \times 10^{-3}\, e\,\bar{g}^{(1)}_{\pi NN}\,/{\rm GeV}\,,
\end{equation}

\item  Taking ${\cal C}^S_{\rm Hg}=-2.8$~\cite{Dzuba:2002kg}
the SIII(HF) coeffcients from Ref.~\cite{Ban:2010ea},
\begin{equation}
\label{eq:dhg4}
d^{\rm \,IV}_{\rm Hg}[S]\ \simeq\
3.1 \times 10^{-4}\, e\,\bar{g}^{(0)}_{\pi NN}\,/{\rm GeV}+
9.5 \times 10^{-5}\, e\,\bar{g}^{(1)}_{\pi NN}\,/{\rm GeV}\,.
\end{equation}

\end{itemize}
Compared  to   the  expression  $d^{\rm  \,I}_{\rm   Hg}[S]$  used  in
Refs.~\cite{Ellis:2008zy,cpsuperh},  $d^{\rm   \,II}_{\rm  Hg}[S]$  is
slightly smaller,  and has a negligible  isoscalar contribution, which
is  actually  ignored  in   $d^{\rm  \,I}_{\rm  Hg}[S]$.   Whilst  the
isoscalar contribution is still  small in $d^{\rm \,III}_{\rm Hg}[S]$,
it can be larger than the  isovector contribution by a factor $\sim 3$
in  $d^{\rm  \,IV}_{\rm Hg}[S]$.  However,  we  observe that  $|d^{\rm
  \,IV}_{\rm Hg}[S]|$  is smaller than $|d^{\rm  \,I}_{\rm Hg}[S]|$ by
about an order of magnitude.

On the other hand, the EDM of $^{225}$Ra is related to its Schiff moment by
\cite{Dzuba:2002kg}
\begin{equation}
d_{\rm Ra}[S]\ =\ -8.5\times 10^{-17}\, e\cdot {\rm cm}
\times \left(\frac{S}{e\cdot {\rm fm}^3}\right)\; .
\end{equation}
Taking the coefficients given in Ref.~\cite{Dobaczewski:2005hz}, see
Table~\ref{tab:Schiff}, we obtain
\begin{eqnarray}
d_{\rm Ra}[S] &\simeq &
-8.7 \times 10^{-2}\, e\,\bar{g}^{(0)}_{\pi NN}\,/{\rm GeV}
+3.5 \times 10^{-1}\, e\,\bar{g}^{(1)}_{\pi NN}\,/{\rm GeV}\,.
\end{eqnarray}
We  note that  the $\bar{g}^{(1)}_{\pi  NN}$ contribution  the  EDM of
$^{225}$Ra is about  200 times larger than to  the Mercury EDM $d^{\rm
  \,I}_{\rm Hg}[S]$, an  enhancement due to the existence  of a nearby
parity-doublet states~\cite{Engel:2003rz}.

\section{Differential-Geometrical Optimization Method}
\label{sec:geo}

We  briefly review our  powerful analytical  approach for  finding the
optimal choice  of CP-odd  phases which maximize  the size of  a given
CP-violating  observable  $O$,  while  remaining compatible  with  the
present EDM constraints~\cite{ELPmath}.  We have applied this approach
previously to estimate  maximal values of the Deuteron  and muon EDMs,
the  CP  asymmetry  in~$b  \to  s  \gamma$, and  the  phase  in  $B_s$
mixing~\cite{ELPouter}, and  it may be  applied similarly to  the case
where the observable $O$ is the EDM of $^{225}$Ra.

We consider  a theory such  as the MCPMFV  SUSY model with  six CP-odd
phases, {\boldmath $\Phi$}, represented  by a 6D phase vector, subject
to three  EDM constraints denoted by $E^{a,b,c}=  0$, corresponding to
the non-observation  of the Thallium,  neutron and Mercury  EDMs.  For
any given value  of the CP-conserving parameters in  the MCPMFV model,
we may  expand these  EDMs and the  observable $O$ in  the small-phase
approximation, defining the four  6D vectors ${\bf E}^{a,b,c} = \nabla
E^{a,b,c}$  and ${\bf  O} =  \nabla O$,  and we  assume that  the four
vectors ${\bf E}^{a,b,c}$ and ${\bf O}$ are linearly independent.

We then introduce the triple exterior product
\begin{equation}
  \label{A3form}
A_{\alpha\beta\gamma}\ \equiv \ E^a_{[\alpha}\,E^b_{\beta}\,E^c_{\gamma ]}\; ,
\end{equation}
where the Greek indices label the  components of the vectors in the 6D
space, i.e., $\alpha ,\ \beta ,\ \gamma\ =\ 1,2,\dots, 6$.  The square
brackets  on  the  RHS  of~(\ref{A3form})  indicate  that  the  tensor
$A_{\alpha\beta\gamma}$  is  obtained  by fully  antisymmetrizing  the
vectors  $E^a_\alpha$,  $E^b_\beta$ and  $E^c_\gamma$  in the  indices
$\alpha   ,   \beta   ,   \gamma$,  i.e.,   $A_{\alpha\beta\gamma}   =
-A_{\beta\alpha\gamma}  = -  A_{\alpha\gamma\beta}$, etc.  Borrowing a
term from the  calculus of differential forms, $A_{\alpha\beta\gamma}$
is a 3-form. We also introduce the 2-form
\begin{equation}
  \label{B2form}
B_{\mu\nu}\ =\ \varepsilon_{\mu\nu\lambda\rho\sigma\tau}\, O_\lambda\,
E^a_\rho\, E^b_\sigma\, E^c_\tau\; ,
\end{equation}
where    summation   over    repeated   indices    is    implied   and
$\varepsilon_{\mu\nu\lambda\rho\sigma\tau}$ is  the usual Levi--Civita
tensor  generalized to  6D.  In  the language  of  differential forms,
$B_{\mu\nu}$ is,  up to an  irrelevant overall factor,  the Hodge-dual
product between the  1-form $O_\lambda$, representing the CP-violating
observable, and the 3-form $A_{\alpha\beta\gamma}$.

The  components  $\Phi^*_\alpha$  of  the optimal  EDM-free  direction
maximizing $O$ can now be  obtained from the Hodge-dual product of the
3-form   $A_{\beta\gamma\delta}$    and   the   2-form   $B_{\mu\nu}$.
Explicitly,
\begin{equation}
  \label{Phi6D}
\Phi^*_\alpha\ =\ {\cal N}\,
\varepsilon_{\alpha\beta\gamma\delta\mu\nu}\,
A_{\beta\gamma\delta}\; B_{\mu\nu}\ =\ 
{\cal N}\; \varepsilon_{\alpha\beta\gamma\delta\mu\nu}\,
\varepsilon_{\mu\nu\lambda\rho\sigma\tau}\,
 E^a_\beta\, E^b_\gamma\, E^c_\delta\; O_\lambda\, E^a_\rho\, E^b_\sigma\, E^c_\tau\ ,
\end{equation}
where we have included  an unknown overall normalization factor ${\cal
  N}$.  By  construction, the 6D phase vector  {\boldmath $\Phi^*$} is
orthogonal  to  the three  vectors  ${\bf  E}^{a,b,c}$, and  therefore
satisfies the desired  EDM constraints, $E^a = E^b = E^c  = 0$, in the
small-phase  approximation.   We observe  that  the magnitude  $\phi^*
\equiv $~{\boldmath  $|\Phi^*|$}, and hence  the overall normalization
factor ${\cal N}$,  can only be determined by  a numerical analysis of
the  actual experimental  limits  on the  three  EDMs.  As  in the  3D
example, the maximum allowed  value of the CP-violating observable $O$
is given in the small-phase approximation by
\begin{equation}
  \label{Ooptimal6}
O \ =\ \phi^*\; \widehat{\Phi}^*_\kappa\,
O_\kappa\,\ =\ \pm\; {\cal N}\ \Big|\varepsilon_{\mu\nu\alpha\beta\gamma\delta}\,
  \varepsilon_{\mu\nu\lambda\rho\sigma\tau}\, O_\alpha\, O_\lambda\,
  E^a_\beta\, E^b_\gamma\, E^c_\delta\, E^a_\rho\, E^b_\sigma\,
  E^c_\tau\,\Big|\;,
\end{equation}
where  the caret  denotes the  components of  a unit-norm  vector.  As
discussed  in~\cite{ELPouter}, quadratic  and  higher-order derivative
terms  with respect  to the  CP-odd phases  will generically  prefer a
particular sign for the optimal value of $O$.

We can  also allow for the  possible presence of a  non-zero strong CP
phase   $\theta_{\rm  QCD}$  in   the  theory,   in  which   case  the
corresponding   CP-odd  phase   vector   {\boldmath  $\Phi$}   becomes
seven-dimensional (7D)  in the MCPMFV SUSY  model.  The generalization
of the above  construction of the optimal value  of the observable $O$
is discussed in Section~5 of~\cite{ELPouter}.

\section{Results}
\label{sec:results}

In this Section we use the above the differential-geometrical approach
to  analyze the  maximal value  of  the $^{225}$Ra  EDM obtainable  in
CP-violating variants of  the following representative CMSSM benchmark
scenario which predicts the mass spectrum of SUSY particles in 
the sub-TeV region:
\begin{eqnarray}
&&\left|M_{1,2,3}\right|=350~~{\rm GeV}\,, \nonumber \\
&&M^2_{H_u}=M^2_{H_d}=\widetilde{M}^2_Q=\widetilde{M}^2_U=\widetilde{M}^2_D
=\widetilde{M}^2_L=\widetilde{M}^2_E=(100~~{\rm GeV})^2\,, \nonumber \\
&&\left|A_u\right|=\left|A_d\right|=\left|A_e\right|=100~~{\rm GeV}\,,
\label{eq:cpsps1a}
\end{eqnarray}
at the GUT scale, introducing non-zero CP-violating phases and varying
$\tan\beta\,(M_{\rm   SUSY})$.    We   adopt   the   convention   that
$\Phi_\mu=0^\circ$, and we vary independently the following six MCPMFV
phases at the GUT scale: $\Phi_1$, $\Phi_2$, $\Phi_{3}$, $\Phi_{A_u}$,
$\Phi_{A_d}$, and $\Phi_{A_e}$.  In the  7D case, in addition to the 6
CP phases, we consider a  non-zero strong CP phase $\theta_{\rm QCD}$.
The scenario (\ref{eq:cpsps1a}) considered in this work is similar
to one with $\left|M_{1,2,3}\right|=250$~GeV that we considered
previously~\cite{ELPouter,Ellis:2008zy,Ellis:2007kb}.
The somewhat larger value $\left|M_{1,2,3}\right|=350$~GeV is chosen
here for consistency with the recent SUSY search results
reported by the CMS Collaboration at the LHC~\cite{Collaboration:2011tk}.
When  $\tan\beta=10$,   $\Phi_{1,2,3}=0^\circ$, and
$\Phi_{A_u,A_d,A_e}=180^\circ$, the previous scenario with $\left|M_{1,2,3}\right|=250$ GeV
became  the well-known  SPS1a point~\cite{SPS} (also known as benchmark B).
The scenario (\ref{eq:cpsps1a}) is more similar to benchmarks C, G and I of~~\cite{SPS}
when $\tan\beta = 10, 20$ or 35, respectively.   Our  calculations  of the  EDMs  are
based   on  Refs.~\cite{ELPouter,Ellis:2008zy},   which   include  the
two-loop  diagrams   mediated  by  the   $\gamma$-$H^\pm$-$W^\mp$  and
$\gamma$-$W^\pm$-$W^\mp$  couplings,  and  we  take into  account  the
effects  of the  different computations  of the  Schiff moment  of the
Mercury nucleus as explained in Section~\ref{sec:schiff}.

\subsection{The MCPMFV SUSY model with 6 CP phases}

In  order to analyze  the scenario  (\ref{eq:cpsps1a}), we  first make
Taylor expansions  of the following four  EDMs in terms  of the MCPMFV
CP-violating phases:
\begin{eqnarray}
&&
d_{\rm Tl}/d_{\rm Tl}^{\rm EXP}\,, \qquad
d_{\rm n}/d_{\rm n}^{\rm EXP}\,, \qquad
d_{\rm Hg}/d_{\rm Hg}^{\rm EXP}\,,  \qquad
d_{\rm Ra}/(10^{-27}\, e\cdot{\rm cm})\,, \ \
\end{eqnarray}
where we choose the following normalization factors:
\begin{equation}
d_{\rm Tl}^{\rm EXP}\ =\ 9\times 10^{-25}\, e\cdot{\rm cm}\,, \qquad
d_{\rm n}^{\rm EXP}\ =\ 3\times 10^{-26}\, e\cdot {\rm cm}\,, \qquad
d_{\rm Hg}^{\rm EXP}\ =\ 3.1\times 10^{-29}\, e\cdot {\rm cm}\,,
\end{equation}
which correspond  to the  current experimental limits  on the  EDMs of
Thallium~\cite{Regan:2002ta},   the  neutron~\cite{Baker:2006ts},  and
Mercury~\cite{Romalis:2000mg,Griffith:2009zz}.     The   normalization
factor for the $^{225}$Ra EDM, namely $10^{-27}\, e\cdot {\rm cm}$, is
typical of the estimated experimental sensitivity.

\begin{figure}[!t]
\begin{center}
{\epsfig{figure=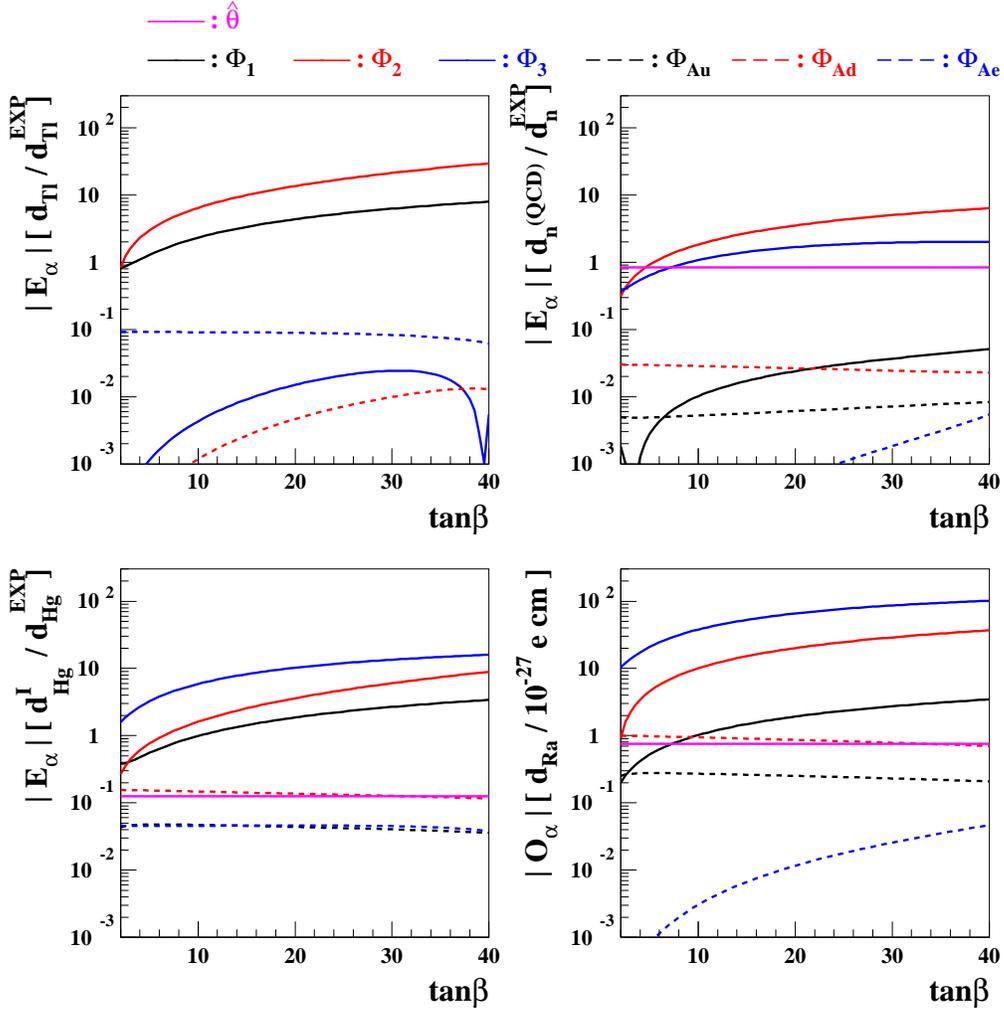,height=14.0cm,width=14.0cm}}
\end{center}
\vspace{-1.0cm}
\caption{\it The absolute values of the components of the three
  vectors ${\bf E}_\alpha$ of MCPMFV CP-violating phases representing
  the present EDM constraints on $d_{\rm Tl}$ (upper left), $d_{\rm
    n}$ (upper right), and $d_{\rm Hg}$ (lower left), and those of the
  vector representing the EDM of $^{225}{\rm Ra}$ (lower right) in
  small-phase expansions around the CP-conserving point, as functions
  of $\tan\beta$ for the scenario (\ref{eq:cpsps1a}).  We use the
  computation employing the QCD sum rule technique and the estimate
  $d_{\rm Hg}^{\rm \,I}$ (\ref{eq:dhg1}) for the neutron and Mercury
  EDMs, respectively.  
The CP-violating phases are measured in degrees.
}
\label{fig:coeff.1}
\end{figure}
\begin{figure}[!t]
\begin{center}
{\epsfig{figure=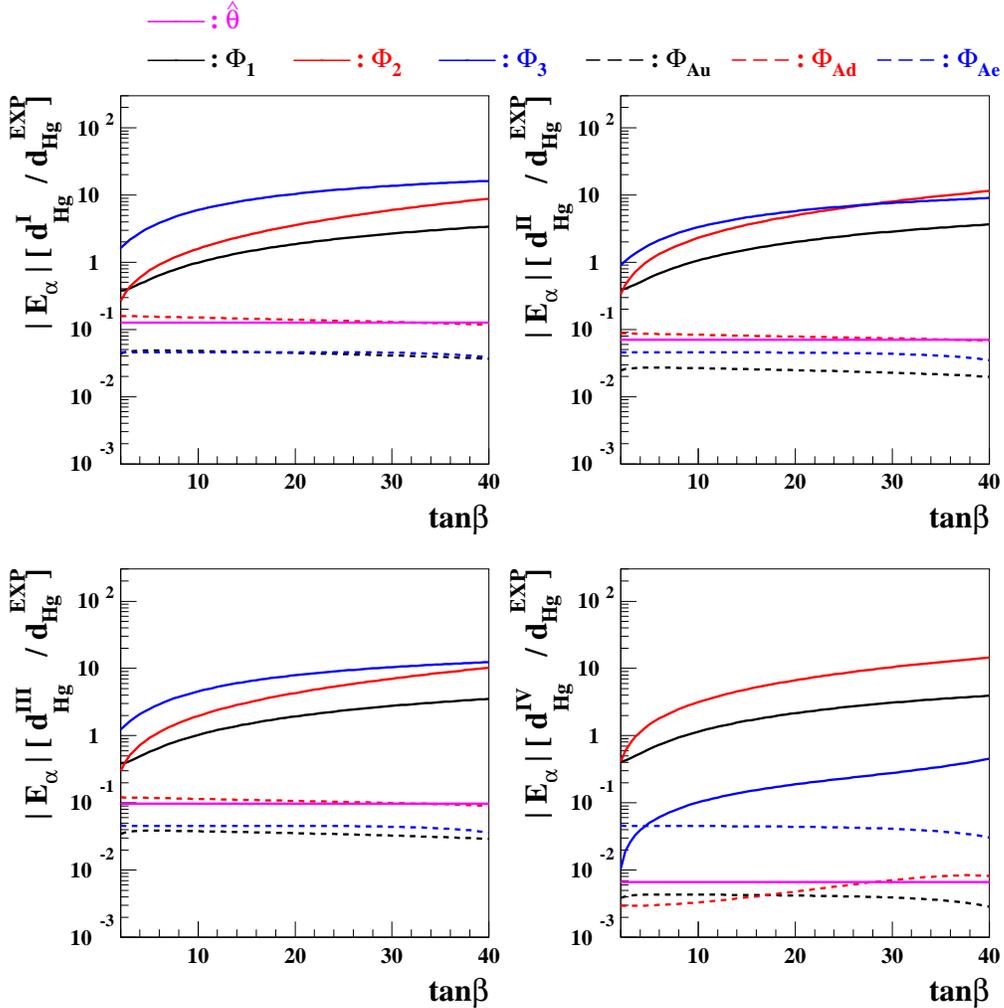,height=14.0cm,width=14.0cm}}
\end{center}
\vspace{-1.0cm}
\caption{\it The absolute values of the components of the
6D vector ${\bf E}_\alpha$ of MCPMFV CP-violating phases 
representing the present EDM constraints on the Mercury EDM,
as obtained using the four different calculations of $d_{\rm Hg}$
discussed in Section~\ref{sec:schiff}:
$d_{\rm Hg}^{\rm \,I}$ (upper left),
$d_{\rm Hg}^{\rm \,II}$ (upper right),
$d_{\rm Hg}^{\rm \,III}$ (lower right), and
$d_{\rm Hg}^{\rm \,IV}$ (lower left). The line styles are the same as
in Fig.~\ref{fig:coeff.1}. 
The CP-violating phases are measured in degrees.
}
\label{fig:coeff.2}
\end{figure}
\begin{figure}[!t]
\begin{center}
{\epsfig{figure=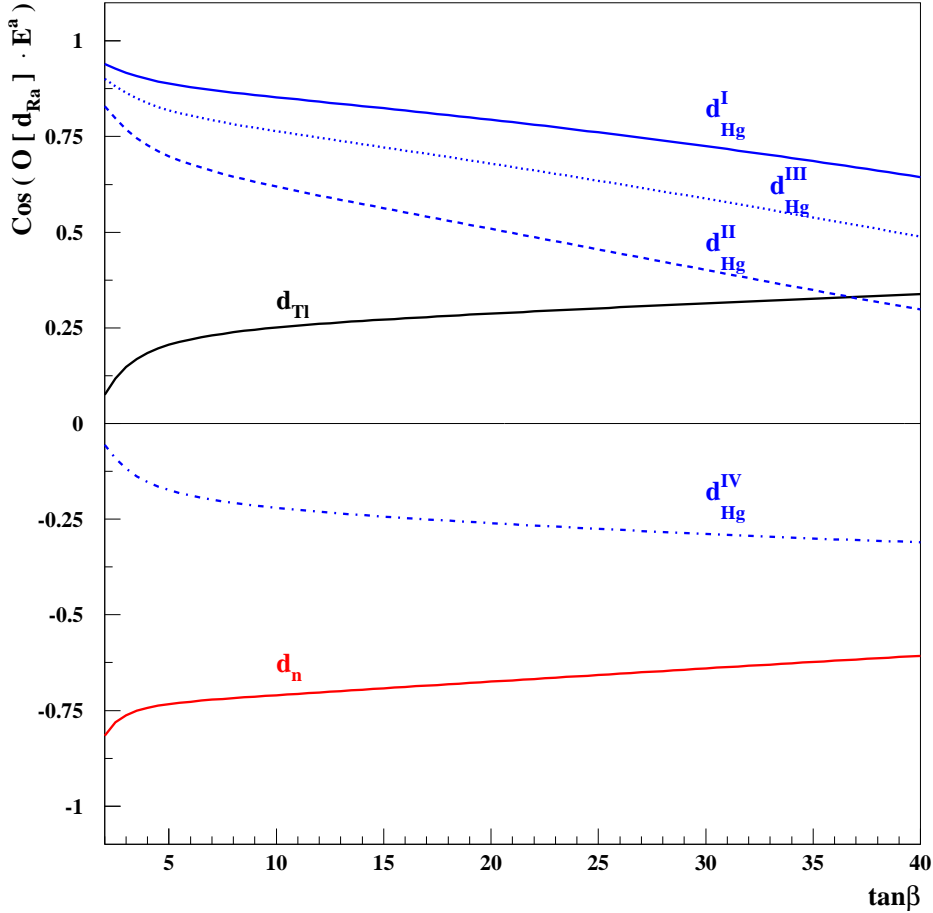,height=14.0cm,width=14.0cm}}
\end{center}
\vspace{-1.0cm}
\caption{\it The  cosines of the angles between the observable 6D vector
${\bf O}^{\,d_{\rm Ra}}$ and the EDM-constraint 6D vectors
${\bf E}^{\,d_{\rm Tl}}$ (black solid),
${\bf E}^{d_{\rm n}}$ (red solid),
${\bf E}^{d_{\rm Hg}^{\rm \,I}}$ (blue solid),
${\bf E}^{d_{\rm Hg}^{\rm \,II}}$ (blue dashed),
${\bf E}^{d_{\rm Hg}^{\rm \,III}}$ (blue dotted), and
${\bf E}^{d_{\rm Hg}^{\rm \,IV}}$ (blue dash-dotted)
as functions of $\tan\beta$.
}
\label{fig:cosine}
\end{figure}
\begin{figure}[!t]
\begin{center}
{\epsfig{figure=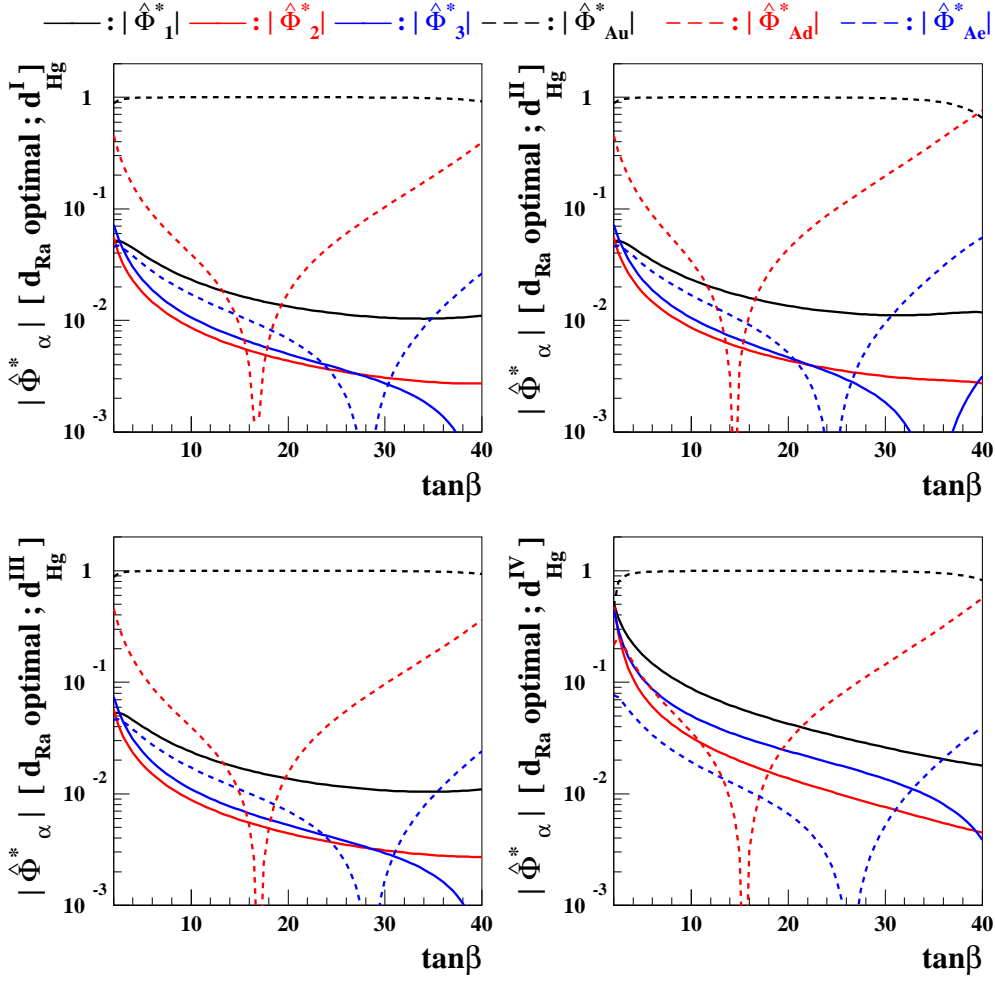,height=14.0cm,width=14.0cm}}
\end{center}
\vspace{-1.0cm}
\caption{\it The absolute values of the six components of
the normalized vectors in the directions where the size
of $d_{\rm Ra}$ is optimized, as obtained using the calculations
$d_{\rm Hg}^{\rm \,I}$ (upper left),
$d_{\rm Hg}^{\rm \,II}$ (upper right),
$d_{\rm Hg}^{\rm \,III}$ (lower right), and
$d_{\rm Hg}^{\rm \,IV}$ (lower left) for the Mercury EDM.
The CP-violating phases are measured in degrees.
}
\label{fig:dir.2}
\end{figure}
\begin{figure}[!t]
\begin{center}
{\epsfig{figure=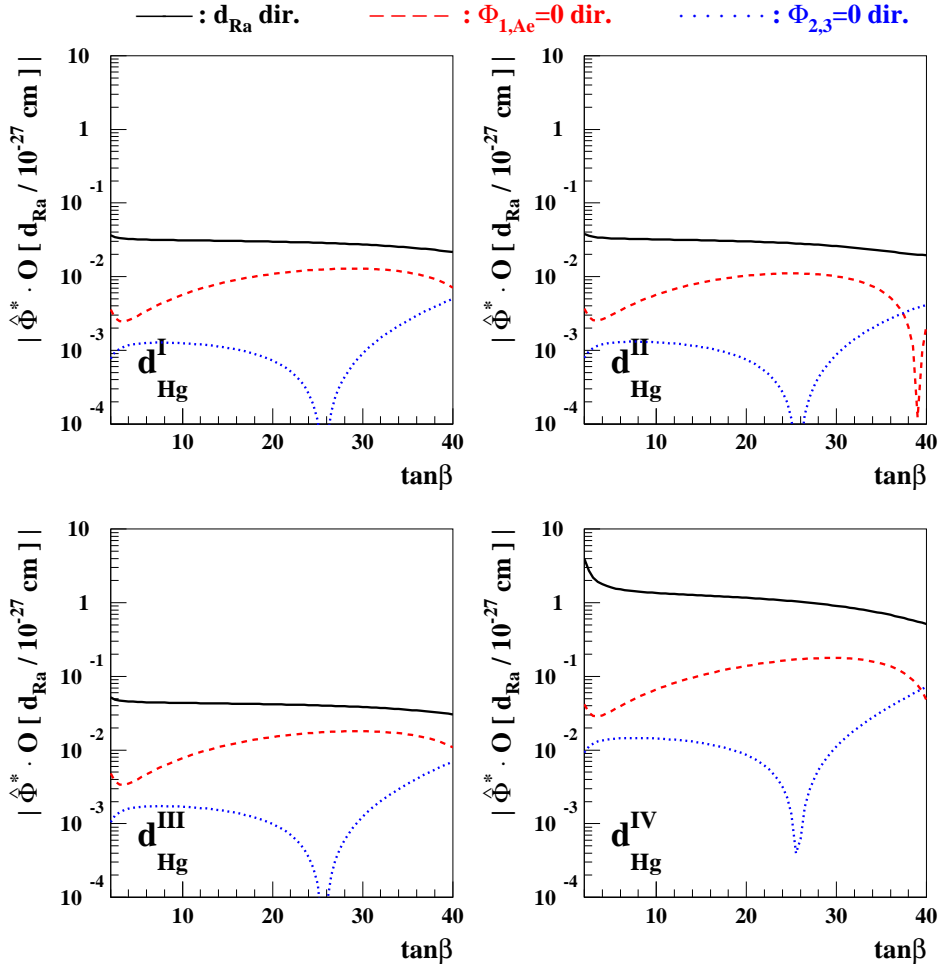,height=14.0cm,width=14.0cm}}
\end{center}
\vspace{-1.0cm}
\caption{\it The products $\widehat{\Phi}^* \cdot {\bf O}$ (black solid)
for the optimal $d_{\rm Ra}$ direction, as obtained using the calculations
$d_{\rm Hg}^{\rm \,I}$ (upper left),
$d_{\rm Hg}^{\rm \,II}$ (upper right),
$d_{\rm Hg}^{\rm \,III}$ (lower right), and
$d_{\rm Hg}^{\rm \,IV}$ (lower left) for the Mercury EDM.
For comparison purposes,  we also show the products along the two
reference directions with
$\Phi_{1,A_e}=0$ (red dashed)
and $\Phi_{2,3}=0$ (blue dotted).
}
\label{fig:coefd}
\end{figure}
\begin{figure}[!t]
\begin{center}
{\epsfig{figure=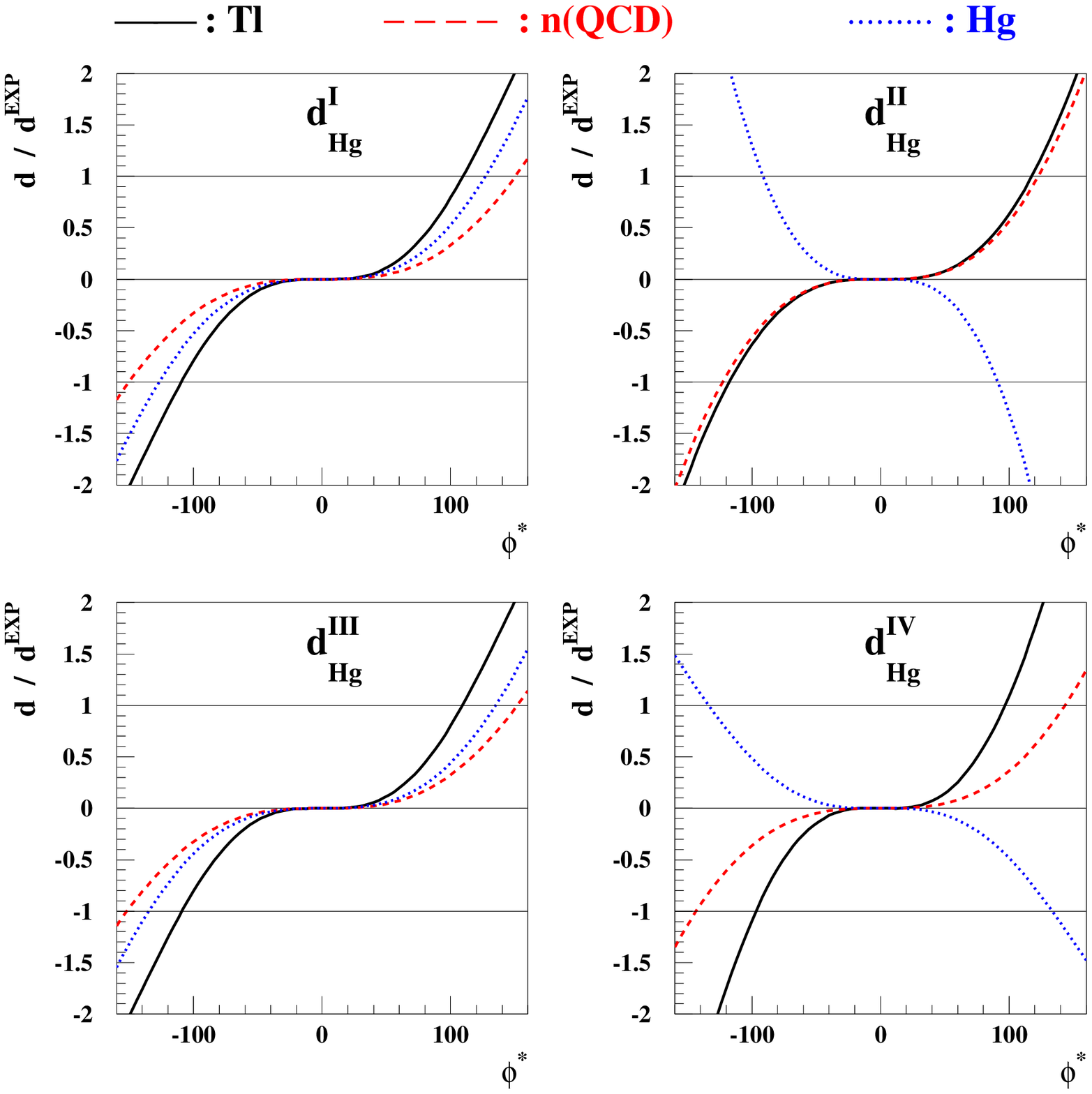,height=14.0cm,width=14.0cm}}
\end{center}
\vspace{-1.0cm}
\caption{\it The values of the three EDMs 
along the directions optimized for $d_{\rm Ra}$, as obtained using
$d_{\rm Hg}^{\rm \,I}$ (upper left),
$d_{\rm Hg}^{\rm \,II}$ (upper right),
$d_{\rm Hg}^{\rm \,III}$ (lower right), and
$d_{\rm Hg}^{\rm \,IV}$ (lower left) for the Mercury EDM.
The ratios
$d_{\rm Tl}/d_{\rm Tl}^{\rm EXP}$,
$d_{\rm n}/d_{\rm n}^{\rm EXP}$, and
$d_{\rm Hg}/d_{\rm Hg}^{\rm EXP}$ are denoted by the
black solid, red dashed, and blue dotted lines, respectively.
The scenario (\ref{eq:cpsps1a}) is assumed, with the choice $\tan\beta=40$.
}
\label{fig:dtlnhg}
\end{figure}
\begin{figure}[!t]
\begin{center}
{\epsfig{figure=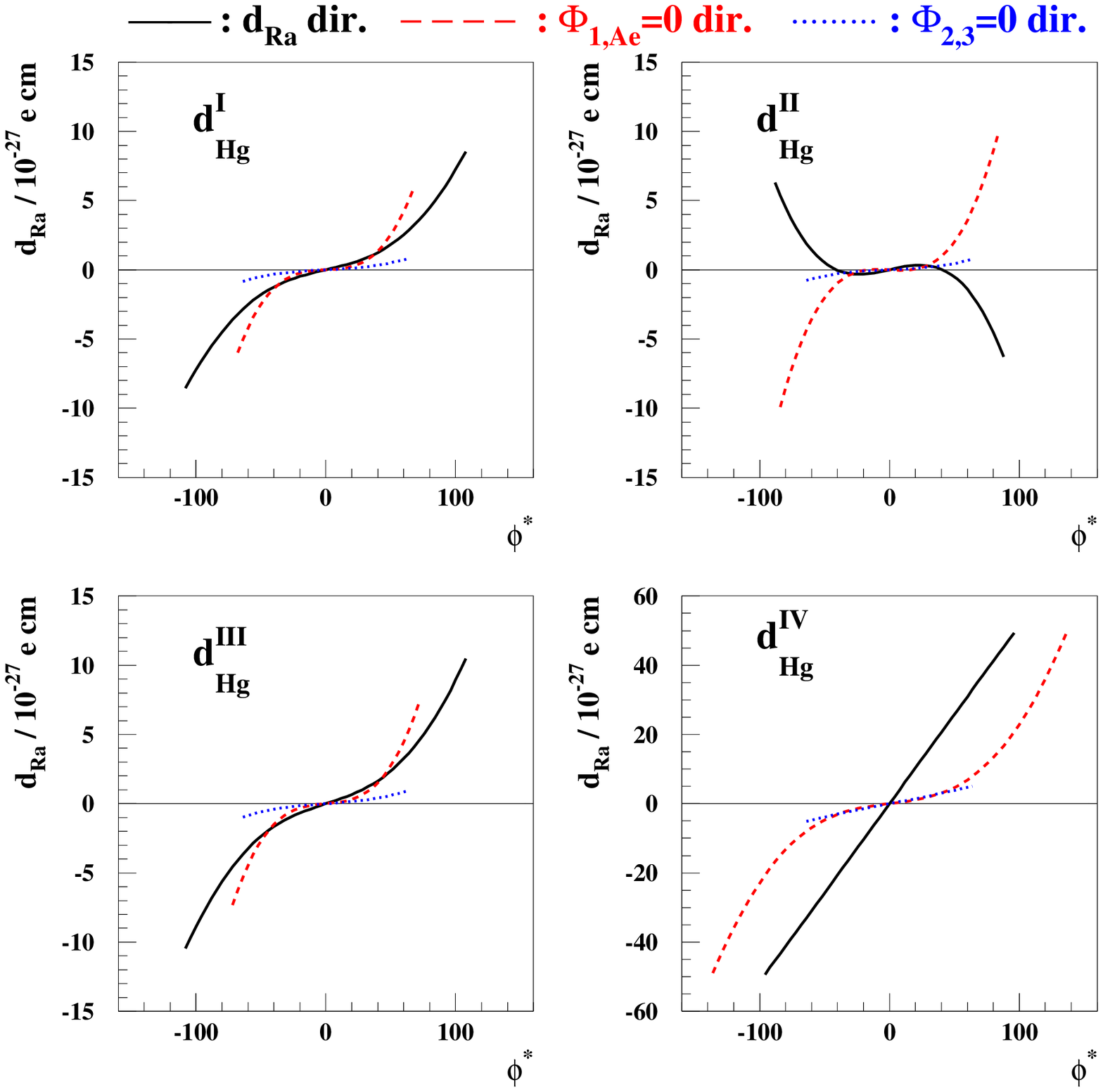,height=14.0cm,width=14.0cm}}
\end{center}
\vspace{-1.0cm}
\caption{\it The EDM of $^{225}$Ra in units of $10^{-27}\, e\,{\rm cm}$ 
for the scenario (\ref{eq:cpsps1a}) with $\tan\beta=40$ 
along the directions optimized for $d_{\rm Ra}$ (black solid)
taking $d_{\rm Hg}^{\rm \,I}$ (upper left),
$d_{\rm Hg}^{\rm \,II}$ (upper right),
$d_{\rm Hg}^{\rm \,III}$ (lower right), and
$d_{\rm Hg}^{\rm \,IV}$ (lower left) for the Mercury EDM.
We have imposed the EDM constraints
$|d_{\rm Tl}/d_{\rm Tl}^{\rm EXP}|\leq 1$,
$|d_{\rm n}/d_{\rm n}^{\rm EXP}|\leq 1$, and
$|d_{\rm Hg}/d_{\rm Hg}^{\rm EXP}|\leq 1$.
For comparison,  we also show $d_{\rm Ra}$ along two reference
directions with the choices
$\Phi_{1,A_e}=0$ (red dashed)
and $\Phi_{2,3}=0$ (blue dotted).
}
\label{fig:dra}
\end{figure}
\begin{figure}[!t]
\begin{center}
{\epsfig{figure=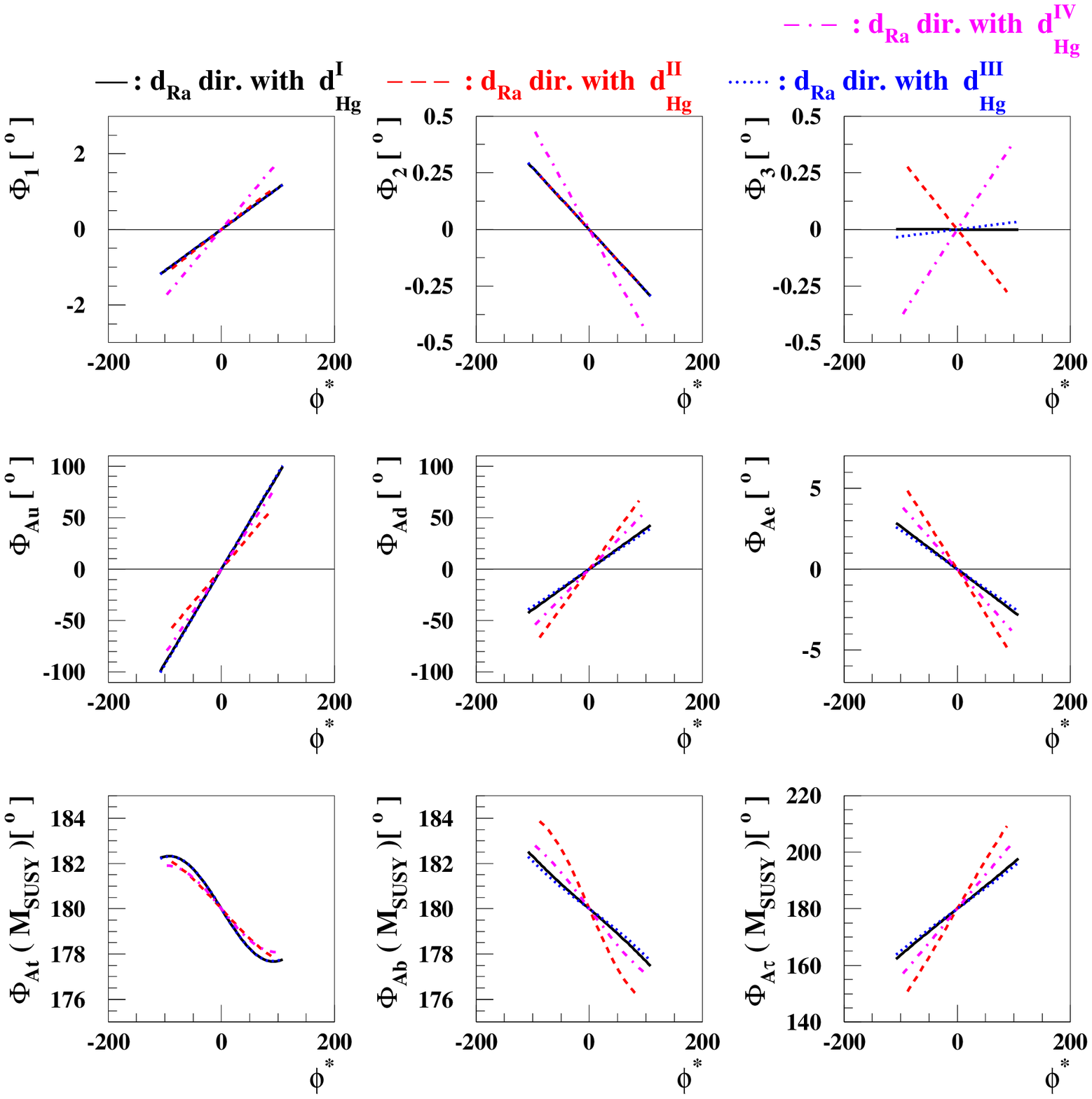,height=14.0cm,width=14.0cm}}
\end{center}
\vspace{-1.0cm}
\caption{\it The 6 CP-violating phases at the GUT scale (upper and
  middle) and the 3 CP-violating phases of the third-generation $A$
  parameters at the SUSY scale (lower) along the directions optimized
  for $d_{\rm Ra}$, as obtained using $d_{\rm Hg}^{\rm \,I}$ (solid
  black), $d_{\rm Hg}^{\rm \,II}$ (red dashed), $d_{\rm Hg}^{\rm
    \,III}$ (blue dotted), and $d_{\rm Hg}^{\rm \,IV}$ (magenta
  dash-dootted) for the Mercury EDM.  We have imposed the EDM
  constraints $|d_{\rm Tl}/d_{\rm Tl}^{\rm EXP}|\leq 1$, $|d_{\rm
    n}/d_{\rm n}^{\rm EXP}|\leq 1$, and $|d_{\rm Hg}/d_{\rm Hg}^{\rm
    EXP}|\leq 1$.  The scenario (\ref{eq:cpsps1a}) is assumed, with
  the choice $\tan\beta=40$.  }
\label{fig:phi}
\end{figure}

In  Fig.~\ref{fig:coeff.1},  we  show   the  absolute  values  of  the
components of the three  6D MCPMFV vectors characterizing the existing
EDM  constraints  and the  6D  vector  representing  the $d_{\rm  Ra}$
observable, for the scenario (\ref{eq:cpsps1a}) varying $\tan\beta$ in
a    small-phase   expansion    around    the   CP-conserving    point
$\Phi_1=\Phi_2=\Phi_3=\Phi_{A_u}=\Phi_{A_d}=\Phi_{A_e}=0^\circ$~\footnote{We
  also  display (as  magenta lines)  the corresponding  components for
  $7$th  components   corresponding  to  the   QCD  phase  ${\overline
    \theta}$, which we discuss later.}.  The solid lines represent the
CP-violating  phases of the  gaugino mass  parameters, and  the dashed
lines  the  trilinear $A$  parameters,  respectively.  The  components
corresponding  to  the CP-violating  gaugino  phases  dominate in  all
cases, increasingly  as $\tan\beta$ grows,  with the exception  of the
$\Phi_3$ component  of ${\bf E}^{\rm d_{\rm Tl}}$.   
For example, when
$\tan\beta=40$, $|d_{\rm  Tl}|$ and $|d_{\rm n}|$ are  larger than the
current  experimental   limits  by  factors   $\sim$30  and  $\sim$5,
respectively,  even when  $\Phi_2=1^\circ$,
whereas  $|d_{\rm Hg}^{\rm \,I}|$  is  larger than  the  current  limit  by 
a  factor  $\sim$15 ($\sim$8) when $\Phi_3=1^\circ$ ($\Phi_2=1^\circ$)
\footnote{
Throughout this work, 
the CP-violating phases are measured in units of degrees.  }.
We see that $|d_{\rm  Ra}|$  can  be as  large  as  $\sim
100\times  10^{-27}\,e\,{\rm  cm}$  ($\sim 40\times  10^{-27}\,e\,{\rm
  cm}$) if $\Phi_3=1^\circ$  ($\Phi_2=1^\circ$).

The relative contributions of the different CP-violating MCPMFV phases
to  $d_{\rm   Hg}$  vary  according  to  the   choice  of  theoretical
calculation,  as  shown  in Fig.~\ref{fig:coeff.2}~\footnote{We again  
display  (as magenta lines)  the corresponding components  for 
possible  $7$th components corresponding  to  the  QCD  
phase ${\overline  \theta}$,  which  we discuss later.}.   
Specifically  we observe the $\Phi_3$ contribution exhibits strong 
variations, and that it is much suppressed if  the 
$|d_{\rm Hg}^{\rm \,IV}|$ calculation is used.     

In Fig.~\ref{fig:cosine} we show the cosines of the angles between the
6D MCPMFV vector representing  the observable ${\bf O}^{\,d_{\rm Ra}}$
and the EDM vectors, which are defined by
\begin{equation}
\label{eq:cosine}
C_{{\bf O,E}^a}\ =\ \frac{ {\bf O}\cdot {\bf E}^a}{|{\bf O}|\, |{\bf E}^a|}\ .
\end{equation}
These  quantities display  the  degrees of  degeneracy, or  alignment,
between the  observable ${\bf O}^{\,d_{\rm  Ra}}$ and the  EDM vectors
$E^a$.  We  see that the  alignments between $d_{\rm Ra}$  and $d_{\rm
  Tl}$ and $d_{\rm Hg}^{\rm \,IV}$ are quite small. On the other hand,
the   alignments   between   $d_{\rm   Ra}$   and   $d_{\rm   Hg}^{\rm
  \,I\,,II\,,III}$  are  larger,   though  decreasing  as  $\tan\beta$
grows.  This  can  be  understood  by observing  that  the  additional
contributions  to the  Mercury EDM,  beyond those  from  the couplings
$\bar{g}^{(0)\,,(1)}_{\pi NN}$, increase with $\tan\beta$.

Having the vectors representing the EDM constraints and the observable
$d_{\rm Ra}$  in hand,  we now combine  them to construct  the optimal
directions   in  the   6D   space  of   CP-violating  MCPMFV   phases,
using~(\ref{Phi6D}),  so as  to maximize  $d_{\rm Ra}$  in  the linear
approximation.   For  comparison,   we  also  consider  two  reference
directions,  which have  $\Phi_1=\Phi_{A_e}=0$  and $\Phi_2=\Phi_3=0$,
respectively.  These  two reference  directions can be  constructed by
defining
\begin{equation}
\Phi^*_\alpha\ \equiv \ {\cal N}\,
\varepsilon_{\alpha\beta\gamma\delta\mu\nu}\,
 E^a_\beta\, E^b_\gamma\, E^c_\delta\; N^{(1)}_\mu\, N^{(2)}_\nu\ ,
\end{equation}
where, for each direction, the two null directions $N^{(1,2)}_\mu$ are
chosen as
\begin{equation}
N^{(1)}_\mu\ =\ (1,0,0,0,0,0)\;,\qquad  N^{(2)}_\mu=(0,0,0,0,0,1)
\label{direction1}
\end{equation}
for the direction $ \Phi_1=\Phi_{A_e}=0$, and
\begin{equation}
N^{(1)}_\mu\ =\ (0,1,0,0,0,0)\;,\qquad  N^{(2)}_\mu\ =\ (0,0,1,0,0,0)
\label{direction2}
\end{equation}
for the direction $\Phi_2=\Phi_3=0$.

We  display in Fig.~\ref{fig:dir.2}   the absolute
values  of the  six  components  of 
the normalized optimal vectors in the direction along which
$d_{\rm Ra}$ is maximized, as obtained using
$d_{\rm Hg}^{\rm \,I}$ (upper left),
$d_{\rm Hg}^{\rm \,II}$ (upper right),
$d_{\rm Hg}^{\rm \,III}$ (lower right), and
$d_{\rm Hg}^{\rm \,IV}$ (lower left) for the Mercury EDM.
We first observe that  the $\Phi_{1,2,3}$ components (solid lines) are
relatively small,  and decrease  as $\tan\beta$ increases.  Hence, all
the  optimal  directions  are  mostly  given by  some  combination  of
$\Phi_{A_u}$ (black  dashed line)  and $\Phi_{A_d}$ (red  dashed line)
directions implying, for $\tan\beta=40$, that $(\Phi_{A_u\,,A_d})^{\rm
  max}  \sim  \phi^*$ whereas  $(\Phi_{1,2,3})^{\rm  max} \sim  \phi^*
\times 10^{-2}$, as will be shown in the following.

In  Fig.~\ref{fig:coefd}, we  consider the  products $\widehat{\Phi}^*
\cdot {\bf O}$ of the  6D vectors in the normalized optimal directions
for  $d_{\rm  Ra}$,  and  the   Radium  EDM,  taking  account  of  the
uncertainty of the Mercury EDM calculation. The products determine the
sizes  of  $d_{\rm  Ra}$  along  its  optimal  direction  through  the
relations given in~(\ref{Ooptimal6})  when $\phi^*=1^\circ$.  As shown
below,  $\phi^*$ could  be  as  large as  $\sim  100^\circ$ before  the
small-phase  approximation  breaks  down  and  one of  the  three  EDM
constraints is  violated.  We  observe that the  direction constructed
using the geometric prescription given in Section~\ref{sec:geo} indeed
gives  the larger values  of $d_{\rm  Ra}$ than  do the  two reference
directions with  $\Phi_{1\,,A_e}=0$ and $\Phi_{2\,,3}=0$.   We note in
particular that,  in the case  of $d_{\rm Hg}^{\rm \,IV}$,  the Radium
EDM may  become about  an order of  magnitude larger than  is possible
with another choices for the Mercury EDM calculation.

Fig.~\ref{fig:dtlnhg} shows the Thallium (solid), neutron (dashed)
and  Mercury (dotted)  EDMs along  the directions  chosen  to optimize
$d_{\rm Ra}$ for the four choices of $d_{\rm Hg}$ when $\tan\beta=40$.
We see that  $\phi^*$ could be as large as  about $100^\circ$, with the
most important  constraints being  provided by the  Thallium (Mercury)
EDM in the cases  where the $d_{\rm Hg}^{\rm \,I\,,III\,,IV}$ ($d_{\rm
  Hg}^{\rm \,II}$) are used, respectively.

In  Fig.~\ref{fig:dra}, we  show the  maximal values  of  $d_{\rm Ra}$
attainable in  the 6D case  after imposing the three  EDM constraints.
We  see  the  that  small-phase approximations  for  $d_{\rm  Hg}^{\rm
  \,I\,,II\,,III}$ break  down for $\phi^* \sim 40^\circ$,  as seen by
comparing with  the calculations  along the two  reference directions,
and  that   $d_{\rm  Ra}$   can  be  as   large  as  $\sim   6  \times
10^{-27}\,e\cdot {\rm cm}$ or more.  On the other hand, in the case of
$d_{\rm  Hg}^{\rm \,IV}$, $d_{\rm  Ra}$ can  be as  large as  $\sim 50
\times 10^{-27}\,e\cdot {\rm cm}$.

Finally, in  Fig.~\ref{fig:phi} we show  the 6 CP-violating  phases at
the GUT  scale (top and middle)  and the 3 CP-violating  phases of the
third-generation $A$ parameters at the SUSY scale (bottom). We observe
that the CP  phases of the gaugino mass  parameters $\Phi_1$, $\Phi_2$
and  $\Phi_3$ can  only  be  as large  as  $2^\circ$, $0.4^\circ$  and
$0.4^\circ$,  respectively, whereas  $\Phi_{A_u}$ ($\Phi_{A_d}$) at
the GUT scale  can be as large as $\sim  100^\circ\,(60^\circ)$, 
as previously seen in Fig.~\ref{fig:dir.2}.  
These  CP-violating phases are suppressed at
the SUSY  scale by RG running from  the GUT scale~\cite{Ellis:2007kb},
but sizeable non-trivial CP-violating  phases are still allowed at the
SUSY     scale:     $|\Phi_{A_t}-180^\circ     |    \sim     2^\circ$,
$|\Phi_{A_b}-180^\circ |  \sim 4^\circ$, and $|\Phi_{A_\tau}-180^\circ
| \sim 30^\circ$.

%-------------------------------------------------------------------
%

\subsection{The 7D Case of non-zero $\theta_{\rm QCD}$}

Hitherto,  we  have  implicitly  assumed  that  the  CP-violating  QCD
$\theta$-term:
\begin{equation}
{\cal L}\ =\ \frac{\alpha_s}{8\pi}\,\bar\theta\,G^a_{\mu\nu}\tilde{G}^{\mu\nu ,a}
\label{eq:theta}
\end{equation}
is     negligible,     where     $\tilde{G}^{\mu\nu     ,a}     \equiv
\epsilon^{\mu\nu\rho\sigma}G^a_{\rho\sigma}/2$   and   the   parameter
$\bar\theta$ is given by the sum of the QCD $\theta_{\rm QCD}$ and the
strong chiral phase for the quark mass matrix:
\begin{equation}
\bar\theta\ =\ \theta_{\rm QCD} +{\rm Arg}\,{\rm Det}\,{M_q}\,.
\end{equation}
In the  weak basis  where ${\rm Arg}\,{\rm  Det}\,{M_q}$ = 0,  we have
$\bar\theta=\theta_{\rm QCD}$.

The   dimension-four  operator   (\ref{eq:theta})  would   in  general
contribute to the neutron, Mercury  and Radium EDMs, e.g., through the
CP-odd     pion-nucleon-nucleon    interactions~(\ref{eq:cpodd_pinn}).
Explicitly,     for     the     neutron     EDM,    we     use     the
estimate~\cite{ELPouter,Pospelov:2005pr}
\begin{eqnarray}
d_{\rm n}(\bar\theta)\  \simeq\ 2.5\times 10^{-16}\,\bar\theta\
e\cdot{\rm cm}.
\end{eqnarray}
For the $\bar\theta$-induced Mercury  EDM, we neglect the contribution
from the $\bar{g}^{(0)}_{\pi NN}$ coupling and use
\begin{equation}
d^{\rm \,I\,,II\,,III\,,IV}_{\rm Hg}(\bar\theta)\ \simeq\ 
(C^{\rm \,I\,,II\,,III\,,IV}_{\rm Hg} \times 10^{-3}\,{\rm GeV}^{-1})
\,e\,\bar{g}^{(1)}_{\pi NN}(\bar\theta)\,,
\end{equation}
where $C^{\rm \,I}_{\rm Hg}=1.8$, $C^{\rm \,II}_{\rm Hg}=1.0$, $C^{\rm
  \,III}_{\rm Hg}=1.4$ and  $C^{\rm \,IV}_{\rm Hg}=9.5\times 10^{-2}$,
with~\cite{ELPouter,Lebedev:2004va}
\begin{equation}
\bar{g}^{(1)}_{\pi NN}(\bar\theta)\
\simeq\ 1.1\times10^{-3}\,\bar\theta\,.
\end{equation}
Finally, for the Radium EDM, we use
\begin{eqnarray}
d_{\rm Ra}(\bar\theta)\  \simeq\ (3.5 \times 10^{-1}\,{\rm GeV}^{-1})
\,e\,\bar{g}^{(1)}_{\pi NN}(\bar\theta)\; .
\end{eqnarray}
Henceforward, we normalize $\bar\theta$ in units of $10^{-10}$:
\begin{equation}
\widehat\theta\ \equiv\ \bar\theta \times 10^{10}\,.
\end{equation}
With this  normalization, when $\widehat\theta  = 1$, we  have $d_{\rm
  n}(\bar\theta)=2.5\times 10^{-26}\, e\cdot  {\rm cm}$, which is very
close to the current  experimental bound $d_{\rm n}^{\rm EXP}=3 \times
10^{-26}\, e\cdot {\rm cm}$, and we find $d_{\rm Ra} \simeq 0.8 \times
10^{-27}\, e\cdot {\rm cm}$.

We have analyzed  the possible maximal values of  $d_{\rm Ra}$ in this
7D case  including $\widehat\theta$  following a procedure  similar to
that we used in the  6D case. Looking again at Fig.~\ref{fig:coeff.1},
we note the horizontal magenta lines representing the $7$th components
of  the vectors representing  the present  EDM constraints  on $d_{\rm
  Tl}$  (upper left),  $d_{\rm  n}$ (upper  right),  and $d_{\rm  Hg}$
(lower left),  and of the  vector representing the EDM  of $^{225}{\rm
  Ra}$  (lower  right).   
%We  observe  that  the  $7$th  component  is
%negligible  in  the $d_{\rm  Tl}$  case,  because  this observable  is
%dominated by the electron EDM. 
We  observe  that  the  $7$th  component  is
missing  in  the $d_{\rm  Tl}$  case,  because  this observable  has no
contribution from the $\widehat\theta$ term in our approach.
On the other hand, the $\widehat\theta$
component is close to unity for $d_{\rm n}$ (cf.~the discussion at the
end of  the previous paragraph),  $\sim 10^{-1}$ for  $d_{\rm Hg}^{\rm
  \,I}$, and somewhat  less than unity for $d_{\rm  Ra}$. This implies
that a measurement of $d_{\rm Ra}$ at the level of $10^{-27}\, e\,{\rm
  cm}$ would already be a competitive measurement of $\widehat\theta$,
even in the absence of the other MCPMFV phases.

As seen in Fig.~\ref{fig:coeff.2}, we have made similar analyses using
the $d_{\rm  Hg}^{\rm \,II\,,III\,,IV}$ calculations,  finding similar
$\widehat\theta$ components in the first  two cases, but a value about
an order of magnitude smaller in the $d_{\rm Hg}^{\rm \,IV}$ case.  We
have also analyzed  (not shown) the cosines of  the angles between the
observable  vector  ${\bf O}^{\,d_{\rm  Ra}}$  and the  EDM-constraint
vectors  ${\bf E}^{\,d_{\rm  Tl}}$,  ${\bf E}^{d_{\rm  n}}$ and  ${\bf
  E}^{d_{\rm  Hg}^{\rm  \,I,\,II,\,III,\,IV}}$  in  the 7D  model,  as
functions of $\tan\beta$. We find results that are very similar to the
6D   case  shown  in   Fig.~\ref{fig:cosine},  the   most  significant
difference being  quite small and limited  to $\tan\beta <  10$ in the
%$d_{\rm Hg}^{\rm \,IV}$ case.
$d_{\rm n}$ case.

\begin{figure}[!t]
\begin{center}
{\epsfig{figure=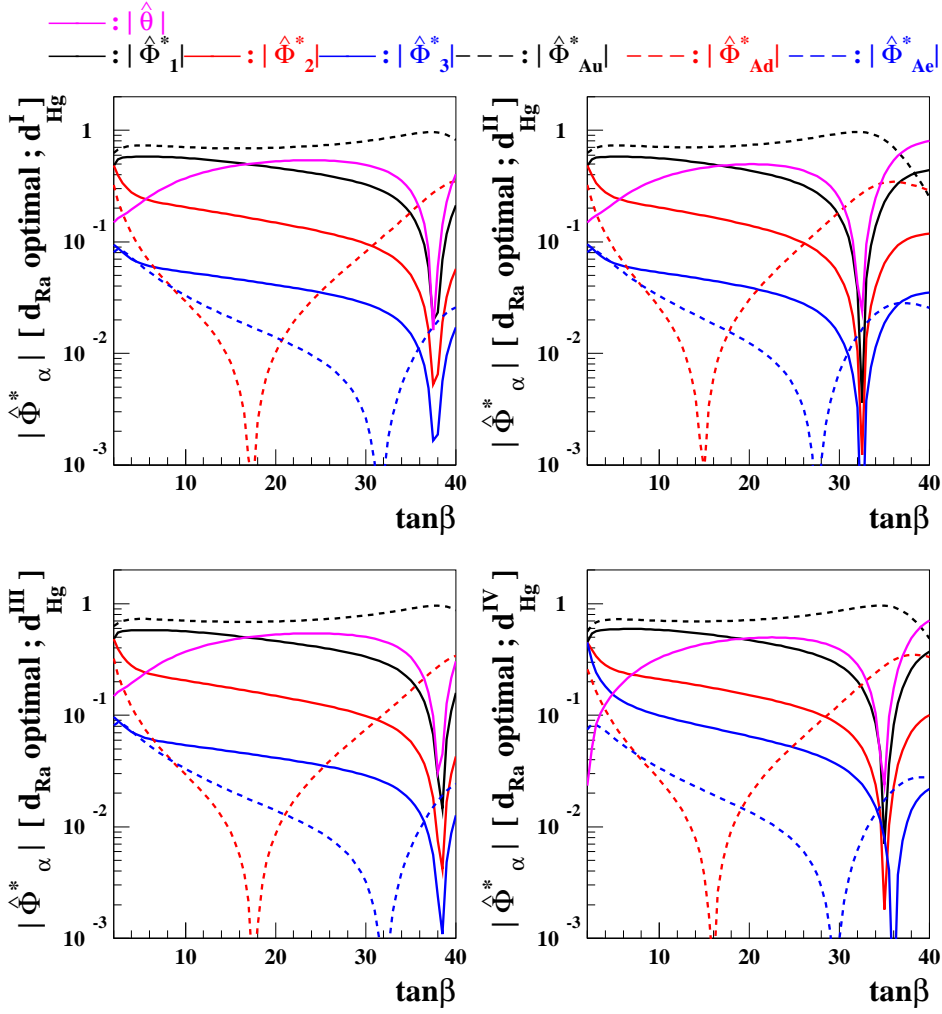,height=14.0cm,width=14.0cm}}
\end{center}
\vspace{-1.0cm}
\caption{\it As in Fig.~\ref{fig:dir.2}, but including the QCD
  $\theta$ term (\protect\ref{eq:theta}). The $7$th components of the
  normalized 7D optimal vectors are denoted by the solid magenta
  lines.  }
\label{fig:dir.2.theta}
\end{figure}

We turn  now to Fig.~\ref{fig:dir.2.theta},  which is the  analogue of
Fig.~\ref{fig:dir.2},   but   including    the   QCD   $\theta$   term
(\ref{eq:theta}).   We  see  significant differences  at  intermediate
and large
$\tan\beta$, where we note that the optimal vectors in all four models
for $d_{\rm  Hg}$ exhibit  not only relatively  large $\widehat\theta$
components,  but also larger  components for  the gaugino  mass phases
than in the  6D case.  We have also analyzed  (not shown) the products
$\widehat{\Phi}^*  \cdot  {\bf  O}$   for  the  optimal  $d_{\rm  Ra}$
direction in the 7D space,  as obtained using the calculations $d_{\rm
  Hg}^{\rm \,I,\,II,\,III,\,IV}$ for  the Mercury EDM, finding results
that    are   generally    very    similar   to    those   shown    in
Fig.~\ref{fig:coefd} with a little rise around $\tan\beta=40$.

\begin{figure}[!t]
\begin{center}
{\epsfig{figure=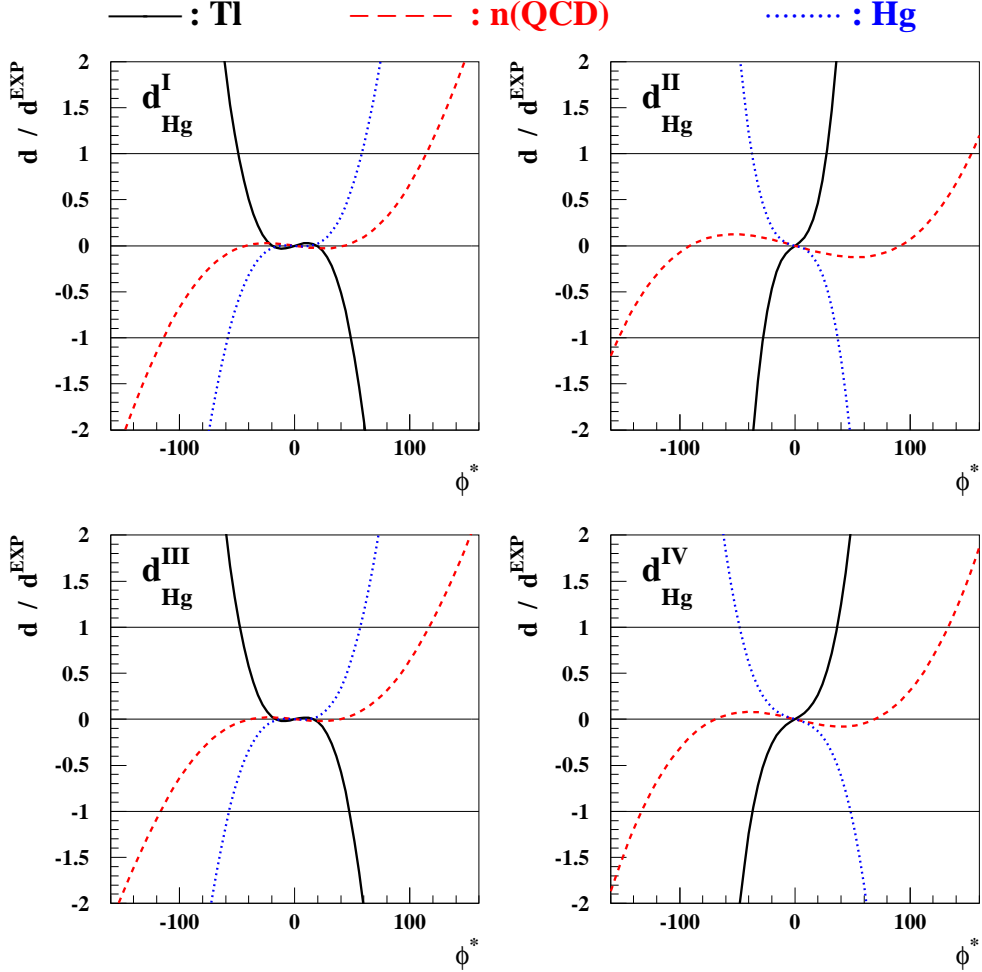,height=14.0cm,width=14.0cm}}
\end{center}
\vspace{-1.0cm}
\caption{\it As in Fig.~\ref{fig:dtlnhg}, but
including the QCD $\theta$ term (\protect\ref{eq:theta}).
}
\label{fig:dtlnhg.theta}
\end{figure}

Figure~\ref{fig:dtlnhg.theta}  displays the  values  of the  Thallium,
neutron and  Mercury EDMs along  the directions optimized  for $d_{\rm
  Ra}$, analogously  to Fig.~\ref{fig:dtlnhg} but this time  in 7D. As
before, the  different panels  display results obtained  using $d_{\rm
  Hg}^{\rm \,I}$ (upper left),  $d_{\rm Hg}^{\rm \,II}$ (upper right),
$d_{\rm Hg}^{\rm  \,III}$ (lower  right), and $d_{\rm  Hg}^{\rm \,IV}$
(lower left)  for the Mercury  EDM, and the ratios  $d_{\rm Tl}/d_{\rm
  Tl}^{\rm  EXP}$,  $d_{\rm   n}/d_{\rm  n}^{\rm  EXP}$,  and  $d_{\rm
  Hg}/d_{\rm  Hg}^{\rm  EXP}$ are  denoted  by  the  black solid,  red
dashed,  and blue  dotted lines,  respectively.  Also  as  before, the
scenario    (\ref{eq:cpsps1a})   is    assumed,   with    the   choice
$\tan\beta=40$.  
%
%Comparing with Fig.~\ref{fig:dtlnhg}, we see that the
%magnitude of the  Thallium EDM is unchanged, but  it sign changes when
%the estimates $d_{\rm Hg}^{\rm \,I,\,III,\,IV}$ are used, but not when
%$d_{\rm Hg}^{\rm \,II}$ is used. The neutron EDM is similar in both 6D
%and 7D  in all  the Mercury calculations,  but the Mercury  EDM itself
%changes sign when the $d_{\rm Hg}^{\rm \,IV}$ calculation is used.
Comparing with Fig.~\ref{fig:dtlnhg}, we see that the
magnitudes of the Thallium and Mercury EDMs increase quickly 
as $\phi^*$ deviates from 0 and 
it could be as large as only about $50^\circ$.

\begin{figure}[!t]
\begin{center}
{\epsfig{figure=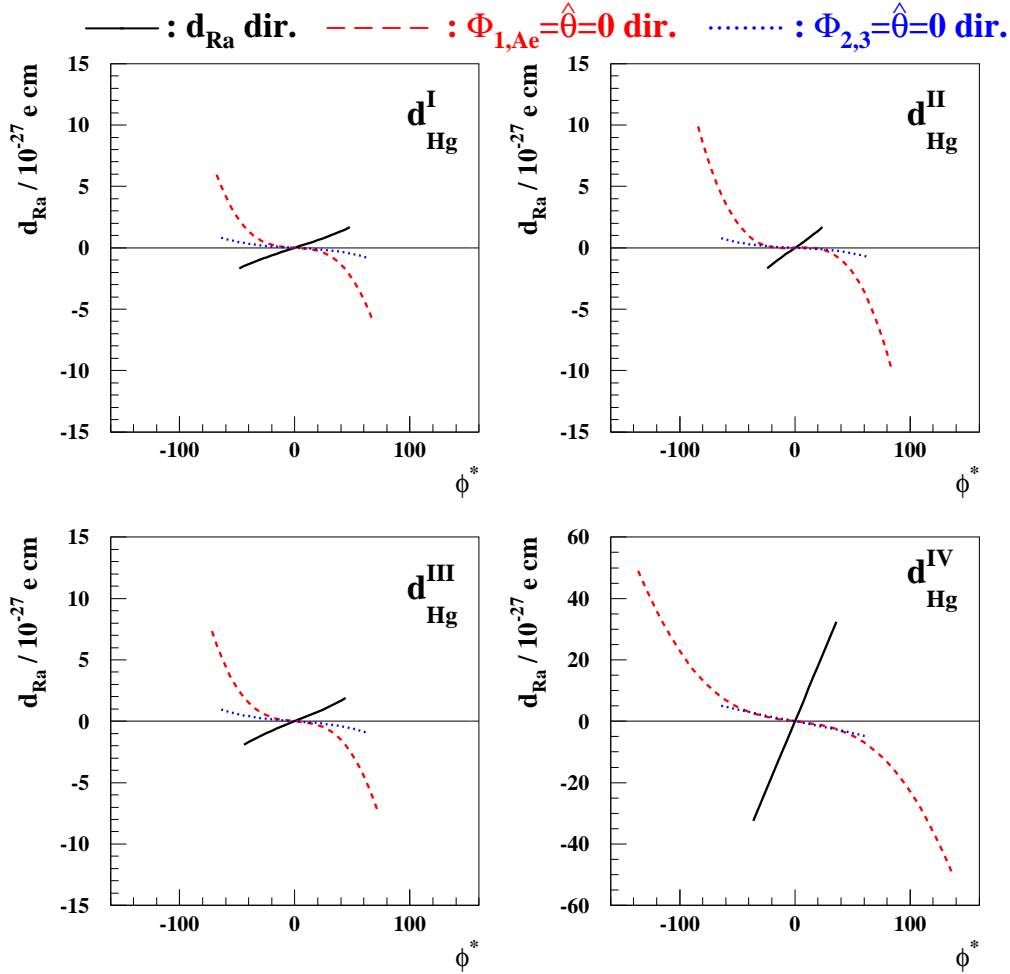,height=14.0cm,width=14.0cm}}
\end{center}
\vspace{-1.0cm}
\caption{\it As in Fig.~\ref{fig:dra}, but
including the QCD $\theta$ term (\protect\ref{eq:theta}).
}
\label{fig:dra.theta}
\end{figure}

\begin{figure}[!t]
\begin{center}
{\epsfig{figure=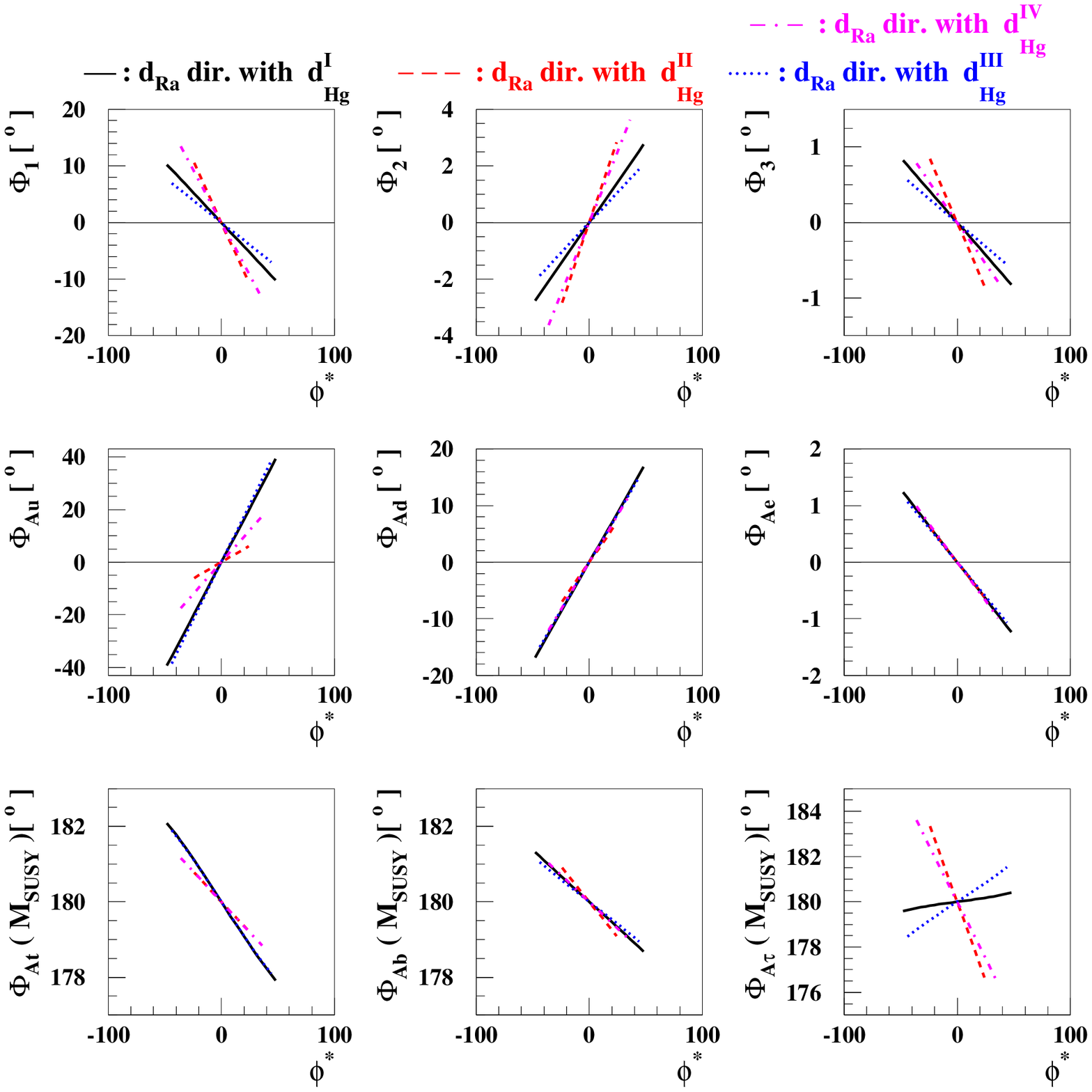,height=14.0cm,width=14.0cm}}
\end{center}
\vspace{-1.0cm}
\caption{\it As in Fig.~\ref{fig:phi}, but
including the QCD $\theta$ term (\protect\ref{eq:theta}).
}
\label{fig:phi.theta}
\end{figure}

\begin{figure}[!t]
\begin{center}
{\epsfig{figure=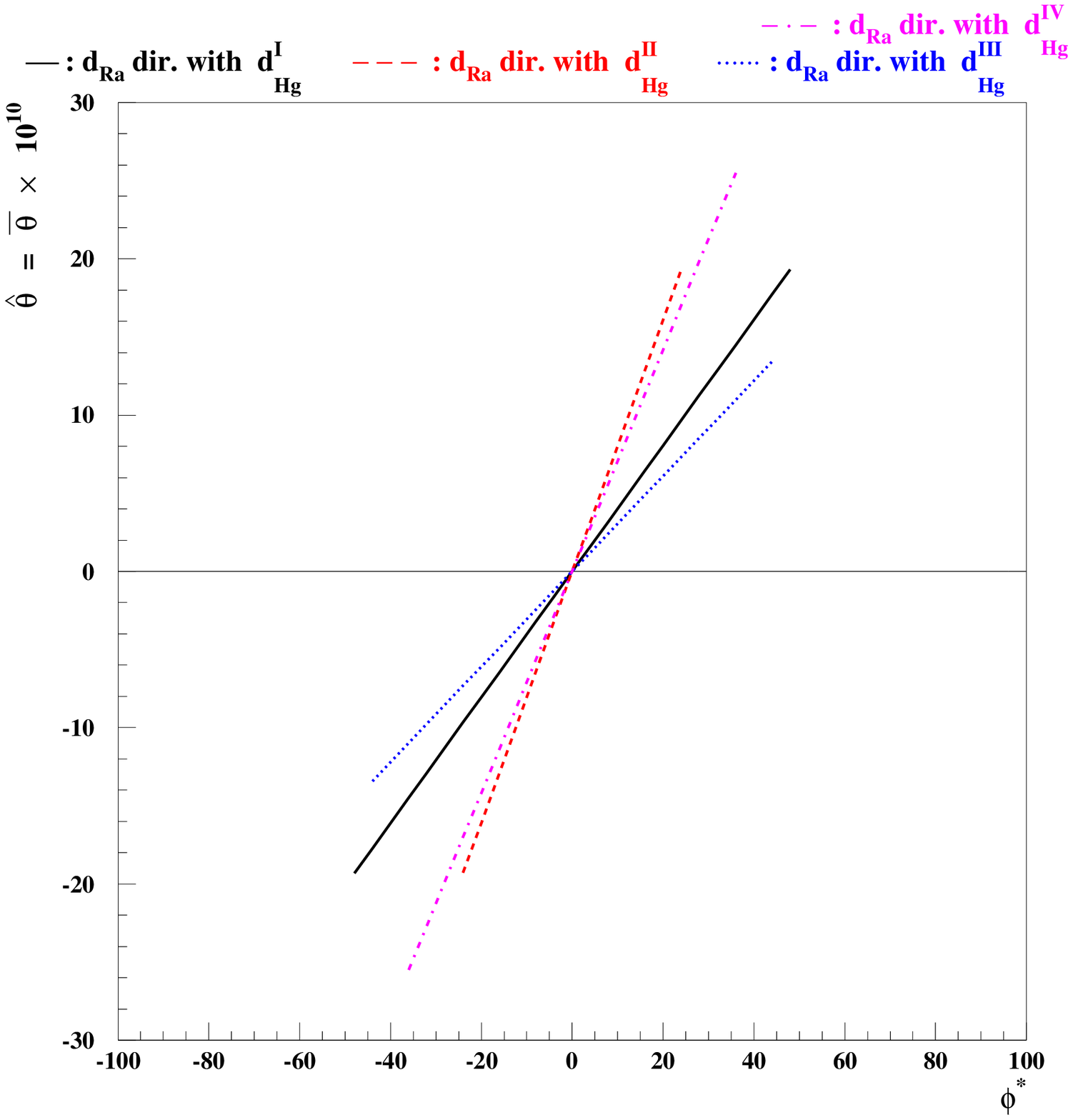,height=13.5cm,width=13.5cm}}
\end{center}
\vspace{-1.0cm}
\caption{\it The values of 
$\hat\theta \equiv \bar\theta \times 10^{10}$
along the directions that maximizer $d_{\rm Ra}$ in the linear approximation,
as found by using the calculations $d_{\rm Hg}^{\rm \,I}$ (solid black),
$d_{\rm Hg}^{\rm \,II}$ (red dashed),
$d_{\rm Hg}^{\rm \,III}$ (blue dotted), and
$d_{\rm Hg}^{\rm \,IV}$ (magenta dash-dootted) for the Mercury EDM.
We have imposed the EDM constraints
$|d_{\rm Tl}/d_{\rm Tl}^{\rm EXP}|\leq 1$,
$|d_{\rm n}/d_{\rm n}^{\rm EXP}|\leq 1$, and
$|d_{\rm Hg}/d_{\rm Hg}^{\rm EXP}|\leq 1$.
The scenario (\ref{eq:cpsps1a}) is assumed, fixing $\tan\beta=40$.
}
\label{fig:tbar.theta}
\end{figure}

Turning  to  the resulting  7D  estimates  of  $d_{\rm Ra}$  shown  in
Fig.~\ref{fig:dra.theta}, 
%we see that  they have similar magnitudes to the 6D estimates 
%shown in  Fig.~\ref{fig:dra} when each of the Mercury
%calculations is  used, though  the $d_{\rm Hg}^{\rm  \,II}$ prediction
%does  change sign.  
we see that  they have somewhat smaller magnitudes than the 6D estimates 
shown in  Fig.~\ref{fig:dra} in the linear approximation.
%In  particular, the  predicted  maximal values  of
%$d_{\rm  Ra}$  are quite  similar  to those  in  the  MCPMFV 6D  case.
However, larger values of  the CP-violating gaugino phases are allowed
in  the  7D   case  than  in  the  6D  case,   as  seen  by  comparing
Figs.~\ref{fig:phi} and  \ref{fig:phi.theta}. Specifically, we observe
that  $\Phi_1$, $\Phi_2$  and  $\Phi_3$  could be  as  large as  $\sim
15^\circ$, $\sim  4^\circ$ and $\sim 1^\circ$,  respectively.  We also
see in  Fig.~\ref{fig:tbar.theta} that sizeable $\overline\theta$ could
be much larger than the  upper limit of $\sim 10^{-10}$ usually quoted,
with  values  as large  as  $\overline\theta  \sim  2.5 \times  10^{-9}$
becoming possible in the presence of non-zero MCPMFV phases.


\section{Conclusions}

In this paper we have extended our previous analyses of the MCPMFV
model, with its 6 CP-violating phases to determine (in the linear
approximation) the largest value of $d_{\rm Ra}$ that is allowed by
the present constraints on the neutron, Thallium and Mercury EDMs,
using the differential-geometric approach developed in~\cite{ELPouter}.
Numerically, we obtain rather similar results whether we include the
CP-violating QCD vacuum phase $\overline\theta$ in the analysis, or not.

The results  are much more  sensitive to the theoretical  treatment of
the Mercury EDM  constraint, and we compare the  results obtained with
four  different calculations  of $d_{\rm  Hg}$. Three  of  them yield
quite similar results for $d_{\rm Ra}$, but one calculations indicates
a smaller dependence of $d_{\rm Hg}$ on the CP-violating phases of the
MCPMFV model, and hence allows  larger numerical values of these phase
and, in general,  larger values of $d_{\rm Ra}$  become possible.  The
maximal values  we find  for $d_{\rm Ra}$  using three of  the $d_{\rm
  Hg}$  calculations are  typically $\sim  6  \times 10^{-27}$~$e\cdot
{\rm cm}$ or more, whereas  the fourth calculation allows $d_{\rm  Ra} \sim 50
\times 10^{-27}$~$e\cdot {\rm cm}$.

For comparison, we recall that  there is a proposal to measure $d_{\rm
  Ra}$  with an sensitivity  approaching $\sim  10^{-27}$~$e\cdot {\rm
  cm}$ in one day of  data-taking.  This experiment would clearly have
interesting potential  to probe regions of the  MCPMFV parameter space
that have not been explored by previous EDM experiments.

This  potential  surely extends  to  many  other  models with  several
sources  of CP  violation,  which  could also  be  analyzed using  the
differential-geometric  approach~\cite{ELPouter}  exploited here.   As
long as a  limited number $n$ of EDMs have  been bounded (or measured)
by experiments, any model with $N > n$ CP-violating parameters will be
underconstrained,  and (partial) cancellations~\cite{IN}  are possible
that would allow large  values for other CP-violating observables. The
MCPMFV  model  is  one such  example,  in  which  $N  =  6$ (or  7  if
$\overline\theta$ is included) and $n =  3$, so far.  As we have shown
in this paper using this differential-geometric approach, constraining
or measuring $d_{\rm Ra}$ at the level of $\sim 10^{-27}$~$e\cdot {\rm
  cm}$ or better would be  a valuable addition to the existing arsenal
of experimental probes of CP violation.

However, it would still not complete the set of constraints needed for
a  model with $N  > 4$,  such as  the MCPMFV.  For this  reason, other
measurements, e.g., of the CP-violating  asymmetry in $b \to s \gamma$
decay.  We  have  discussed  elsewhere  the maximal  value  that  this
observable might  take in the MCPMFV,  and it would  be an interesting
and complementary measurement to that of $d_{\rm Ra}$.

We conclude by drawing reader's  attention to potential caveats in the
physics of CP violation.  The  baryon asymmetry of the Universe is one
of the  strongest pieces of  evidence for physics beyond  the Standard
Model,  as it  cannot be  generated successfully  within  the standard
Kobayashi--Maskawa model  of CP violation.  There must  be new sources
of CP  violation beyond the Kobayashi--Maskawa phase,  and it behooves
experiments to  chase down all those within  reach.  Some CP-violating
phases may manifest  themselves at the TeV scale  and be accessible to
contemporary   collider  experiments,   e.g.~at  the   LHC.   However,
baryogenesis could  equally well  be achieved via  CP-violating phases
appearing  at higher  energy scales,  and EDMs  have the  potential to
probe beyond the  TeV scale, in particular because  the Standard Model
Kobayashi--Maskawa  predictions  for EDMs  are  quite  small.  As  our
analysis exemplifies,  new EDM observables  probe complementary region
of parameter  space including the strong CP  phase $\theta_{\rm QCD}$.
Therefore,  constraining   or  even  measuring  $d_{\rm   Ra}$  is  an
interesting experimental objective.

\subsection*{Acknowledgements}

We thank Jon Engel, Maxim Pospelov and Adam Ritz for useful comments
and suggestions.

%
%\begin{figure}[!t]
%\begin{center}
%{\epsfig{figure=coeff.1.theta.eps,height=14.0cm,width=14.0cm}}
%\end{center}
%\vspace{-1.0cm}
%\caption{\it  As in Fig.~\ref{fig:coeff.1}, but
%including the QCD $\theta$ term (\protect\ref{eq:theta}).
%The solid magneta lines represent the $7$th components
%of the three 7D vectors representing the present EDM constraints on
%$d_{\rm Tl}$ (upper left), $d_{\rm n}$ (upper right), and
%$d_{\rm Hg}$ (lower left), and of the 7D vector
%representing the EDM of $^{225}{\rm Ra}$ (lower right).
%}
%\label{fig:coeff.1.theta}
%\end{figure}
%
%\begin{figure}[!t]
%\begin{center}
%{\epsfig{figure=coeff.2.theta.eps,height=14.0cm,width=14.0cm}}
%\end{center}
%\vspace{-1.0cm}
%\caption{\it As in Fig.~\ref{fig:coeff.2}, but
%including the QCD $\theta$ term (\protect\ref{eq:theta}). The solid magenta lines represent 
%the $7$th components of the four 7D EDM vectors.
%}
%\label{fig:coeff.2.theta}
%\end{figure}
%
%\begin{figure}[!t]
%\begin{center}
%{\epsfig{figure=cosine.theta.eps,height=14.0cm,width=14.0cm}}
%\end{center}
%\vspace{-1.0cm}
%\caption{\it As in Fig.~\ref{fig:cosine}, but
%including the QCD $\theta$ term (\protect\ref{eq:theta}). 
%}
%\label{fig:cosine.theta}
%\end{figure}
%
%
%\begin{figure}[!t]
%\begin{center}
%{\epsfig{figure=coefd.theta.eps,height=14.0cm,width=14.0cm}}
%\end{center}
%\vspace{-1.0cm}
%\caption{\it As in Fig.~\ref{fig:coefd}, but
%including the QCD $\theta$ term (\protect\ref{eq:theta}).
%}
%\label{fig:coefd.theta}
%\end{figure}
%


\newpage
\begin{thebibliography}{99}

\bibitem{nath} For recent reviews, see,\\
T.~Ibrahim and P.~Nath,
``CP violation from standard model to strings,''
  Rev.\ Mod.\ Phys.\  {\bf 80} (2008) 577;\\
%  [arXiv:0705.2008 [hep-ph]].
%\cite{Nath:2010zj}
%\bibitem{Nath:2010zj}
P.~Nath {\it et al.},
``The Hunt for New Physics at the Large Hadron Collider,''
  Nucl.\ Phys.\ Proc.\ Suppl.\  {\bf 200-202} (2010) 185.
%  [arXiv:1001.2693 [hep-ph]].
  %%CITATION = NUPHZ,200-202,185;%%

%\cite{Ellis:2007kb}
\bibitem{Ellis:2007kb}
  J.~R.~Ellis, J.~S.~Lee and A.~Pilaftsis,
  ``B-Meson Observables in the Maximally CP-Violating MSSM with Minimal Flavour
  Violation,''
  Phys.\ Rev.\  D {\bf 76} (2007) 115011.
%  [arXiv:0708.2079 [hep-ph]].
  %%CITATION = PHRVA,D76,115011;%%

\bibitem{RefMFV} For related approaches, see,\\
  M.~Argyrou, A.~B.~Lahanas and V.~C.~Spanos,
  ``Refining the predictions of supersymmetric CP-violating models: A top-down
  approach,''
  JHEP {\bf 0805} (2008) 026;\\
  %[arXiv:0804.2613 [hep-ph]].
  G~Colangelo, E.~Nikolidakis and C.~Smith,
  ``Supersymmetric models with minimal flavour violation and their running,''
  Eur.\ Phys.\ J.\  C {\bf 59} (2009) 75;\\
  %[arXiv:0807.0801 [hep-ph]]
  W.~Altmannshofer, A.~J.~Buras and P.~Paradisi,
  ``Low Energy Probes of CP Violation in a Flavor Blind MSSM,''
  Phys.\ Lett.\  B {\bf 669} (2008) 239;\\
  %[arXiv:0808.0707 [hep-ph]].
  L.~Mercolli and C.~Smith,
  ``EDM constraints on flavored CP-violating phases,''
  Nucl.\ Phys.\  B {\bf 817} (2009) 1; \\
  A.~L.~Kagan, G.~Perez, T.~Volansky and J.~Zupan,
  ``General Minimal Flavor Violation,''
  Phys.\ Rev.\  D {\bf 80} (2009) 076002; \\
  %[arXiv:0903.1794 [hep-ph]].
  R.~Zwicky and T.~Fischbacher,
  ``On discrete Minimal Flavour Violation,''
  Phys.\ Rev.\  D {\bf 80} (2009) 076009.
  %[arXiv:0908.4182 [hep-ph]].
  %%CITATION = PHRVA,D80,076009;%%

\bibitem{IN}
T.~Ibrahim and P.~Nath,
``The neutron and the lepton EDMs in MSSM, large CP violating phases, and
the cancellation mechanism,''
Phys.\ Rev.\  D {\bf 58} (1998) 111301
[Erratum-ibid.\  D {\bf 60} (1999) 099902]; \\
%[arXiv:hep-ph/9807501].
%%%CITATION = PHRVA,D58,111301;%%
M.~Brhlik, L.~L.~Everett, G.~L.~Kane and J.~D.~Lykken,
``A resolution to the supersymmetric CP problem with large soft phases  via
D-branes,''
Phys.\ Rev.\ Lett.\  {\bf 83} (1999) 2124.
%[arXiv:hep-ph/9905215].
%%CITATION = PRLTA,83,2124;%%

\bibitem{ELPouter}
%\cite{Ellis:2010xm}
%\bibitem{Ellis:2010xm}
J.~Ellis, J.~S.~Lee and A.~Pilaftsis,
  ``A Geometric Approach to CP Violation: Applications to the MCPMFV SUSY
  Model,''
JHEP {\bf 1010} (2010) 049.
%  [arXiv:1006.3087 [hep-ph]].
  %%CITATION = JHEPA,1010,049;%%

\bibitem{ELPmath}
%\cite{Ellis:2010uu}
%\bibitem{Ellis:2010uu}
  J.~Ellis, J.~S.~Lee and A.~Pilaftsis,
  ``Note on a Differential-Geometrical Construction of Optimal Directions in
  Linearly-Constrained Systems,''
  arXiv:1009.1151 [math.OC].
  %%CITATION = ARXIV:1009.1151;%%

\bibitem{Schiff} L.~I. Schiff, Phys.~Rev.~{\bf 132} (1963) 2194.

\bibitem{Willmann}
L. Willmann, K. Jungmann, H. W. Wilschut, ``Searches for permanent 
electric dipole moments in Radium Isotopes'', 
Letter of Intent to the ISOLDE and Neutron Time-of-Flight Experiments Committee
for experiments with HIE-ISOLDE, CERN-INTC-2010-049 / INTC-I-115.

\bibitem{Butler}
J. Pakarinen {\it et al.}, ``Measurements of octupole collectivity in 
odd-mass Rn, Fr and Ra isotopes'', Letter of Intent to the ISOLDE and Neutron 
Time-of-Flight Experiments Committee for experiments with HIE-ISOLDE, 
CERN-INTC-2010-022 / INTC-I-091.

%\cite{Khriplovich:1999qr}
\bibitem{Khriplovich:1999qr}
  I.~B.~Khriplovich and R.~A.~Korkin,
  ``P and T odd electromagnetic moments of deuteron in chiral limit,''
  Nucl.\ Phys.\  A {\bf 665} (2000) 365.
% [arXiv:nucl-th/9904081].
  %%CITATION = NUPHA,A665,365;%%

%\cite{Pospelov:2005pr}
\bibitem{Pospelov:2005pr}
  M.~Pospelov and A.~Ritz,
  ``Electric dipole moments as probes of new physics,''
  Annals Phys.\  {\bf 318} (2005) 119.
%  [arXiv:hep-ph/0504231].
  %%CITATION = APNYA,318,119;%%

%\cite{Ban:2010ea}
\bibitem{Ban:2010ea}
  S.~Ban, J.~Dobaczewski, J.~Engel and A.~Shukla,
  ``Fully self-consistent calculations of nuclear Schiff moments,''
  arXiv:1003.2598 [nucl-th].
  %%CITATION = ARXIV:1003.2598;%%

%\cite{Dmitriev:2003kb}
\bibitem{Dmitriev:2003kb}
  V.~F.~Dmitriev and R.~A.~Sen'kov,
  ``P- and T-violating Schiff moment of the Mercury nucleus,''
  Phys.\ Atom.\ Nucl.\  {\bf 66}, 1940 (2003)
  [Yad.\ Fiz.\  {\bf 66}, 1988 (2003)].
%  [arXiv:nucl-th/0304048].
  %%CITATION = YAFIA,66,1988;%%

%\cite{Ellis:2008zy}
\bibitem{Ellis:2008zy}
  J.~R.~Ellis, J.~S.~Lee and A.~Pilaftsis,
  ``Electric Dipole Moments in the MSSM Reloaded,''
  JHEP {\bf 0810} (2008) 049.
%  [arXiv:0808.1819 [hep-ph]].
  %%CITATION = JHEPA,0810,049;%%

\bibitem{cpsuperh}
%\cite{Lee:2003nt}
%\bibitem{Lee:2003nt}
  J.~S.~Lee, A.~Pilaftsis, M.~Carena, S.~Y.~Choi, M.~Drees, J.~R.~Ellis and
C.~E.~M.~Wagner,  ``CPsuperH: A computational tool for Higgs
phenomenology in the 
minimal   supersymmetric standard model with explicit CP violation,''
  Comput.\ Phys.\ Commun.\  {\bf 156} (2004) 283;\\
%  [arXiv:hep-ph/0307377];
  %%CITATION = HEP-PH 0307377;%%
%\cite{Lee:2007gn}
%\bibitem{Lee:2007gn}
  J.~S.~Lee, M.~Carena, J.~Ellis, A.~Pilaftsis and C.~E.~M.~Wagner,
  ``CPsuperH2.0: an Improved Computational Tool for Higgs Phenomenology in the
  MSSM with Explicit CP Violation,''
Comput.\ Phys.\ Commun.\  {\bf 180} (2009) 312.
%  [arXiv:0712.2360 [hep-ph]].
  %%CITATION = CPHCB,180,312;%%

%\cite{deJesus:2005nb}
\bibitem{deJesus:2005nb}
  J.~H.~de Jesus and J.~Engel,
  ``Time-Reversal-Violating Schiff Moment of 199Hg,''
  Phys.\ Rev.\  C {\bf 72} (2005) 045503.
%  [arXiv:nucl-th/0507031].
  %%CITATION = PHRVA,C72,045503;%%

%\cite{Engel:2003rz}
\bibitem{Engel:2003rz}
  J.~Engel, M.~Bender, J.~Dobaczewski, J.~H.~De Jesus and P.~Olbratowski,
  ``Time-Reversal Violating Schiff Moment of 225Ra,''
  Phys.\ Rev.\  C {\bf 68}, 025501 (2003).
%  [arXiv:nucl-th/0304075].
  %%CITATION = PHRVA,C68,025501;%%

%\cite{Dobaczewski:2005hz}
\bibitem{Dobaczewski:2005hz}
  J.~Dobaczewski and J.~Engel,
  ``Nuclear time-reversal violation and the Schiff moment of 225Ra,''
  Phys.\ Rev.\ Lett.\  {\bf 94}, 232502 (2005).
%  [arXiv:nucl-th/0503057].
  %%CITATION = PRLTA,94,232502;%%

%\cite{Dmitriev:2004fk}
\bibitem{Dmitriev:2004fk}
  V.~F.~Dmitriev, R.~A.~Sen'kov and N.~Auerbach,
  ``Effects of core polarization on the nuclear Schiff moment,''
  Phys.\ Rev.\  C {\bf 71} (2005) 035501.
%  [arXiv:nucl-th/0408065].
  %%CITATION = PHRVA,C71,035501;%%

%\cite{Pospelov:2001ys}
\bibitem{Pospelov:2001ys}
  M.~Pospelov,
  ``Best values for the CP-odd meson nucleon couplings from supersymmetry,''
  Phys.\ Lett.\  B {\bf 530} (2002) 123.
%  [arXiv:hep-ph/0109044].
  %%CITATION = PHLTA,B530,123;%%

\bibitem{private:PospelovRitz}
M.~Pospelov and A.~Ritz, in private communication.

%\cite{Lebedev:2002ne}
\bibitem{Lebedev:2002ne}
  O.~Lebedev and M.~Pospelov,
  ``Electric dipole moments in the limit of heavy superpartners,''
  Phys.\ Rev.\ Lett.\  {\bf 89} (2002) 101801
  [arXiv:hep-ph/0204359].
  %%CITATION = PRLTA,89,101801;%%

%\cite{Demir:2003js}
\bibitem{Demir:2003js}
  D.~A.~Demir, O.~Lebedev, K.~A.~Olive, M.~Pospelov and A.~Ritz,
  ``Electric dipole moments in the MSSM at large $\tan\beta$,''
  Nucl.\ Phys.\  B {\bf 680} (2004) 339.
%  [arXiv:hep-ph/0311314].
  %%CITATION = NUPHA,B680,339;%%

%\cite{Romalis:2000mg}
\bibitem{Romalis:2000mg}
  M.~V.~Romalis, W.~C.~Griffith and E.~N.~Fortson,
  ``A new limit on the permanent electric dipole moment of Hg-199,''
  Phys.\ Rev.\ Lett.\  {\bf 86} (2001) 2505.
  %[arXiv:hep-ex/0012001].
  %%CITATION = PRLTA,86,2505;%%

%\cite{Griffith:2009zz}
\bibitem{Griffith:2009zz}
W.~C.~Griffith, M.~D.~Swallows, T.~H.~Loftus, M.~V.~Romalis,
B.~R.~Heckel and E.~N.~Fortson,
  ``Improved Limit on the Permanent Electric Dipole Moment of Hg-199,''
  Phys.\ Rev.\ Lett.\  {\bf 102} (2009) 101601.
%%CITATION = PRLTA,102,101601;%%

%\bibitem{XeEDM}
%M.~A.~Rosenberry and T.~E.~Chupp,
%``Atomic Electric Dipole Moment Measurement Using Spin
%Exchange Pumped Masers of 129Xe and 3He,''
%Phys.\ Rev.\ Lett.\  {\bf 86} (2001) 22.

%\cite{Flambaum:1985ty}
\bibitem{Flambaum:1985ty}
  V.~V.~Flambaum, I.~B.~Khriplovich and O.~P.~Sushkov,
  ``Limit On The Constant Of T Nonconserving Nucleon Nucleon Interaction,''
  Phys.\ Lett.\  B {\bf 162}, 213 (1985).
  %%CITATION = PHLTA,B162,213;%%

%\cite{Flambaum:1985gv}
\bibitem{Flambaum:1985gv}
  V.~V.~Flambaum, I.~B.~Khriplovich and O.~P.~Sushkov,
  ``On The P And T Nonconserving Nuclear Moments,''
  Nucl.\ Phys.\  A {\bf 449}, 750 (1986).
  %%CITATION = NUPHA,A449,750;%%

%\cite{Dzuba:2002kg}
\bibitem{Dzuba:2002kg}
  V.~A.~Dzuba, V.~V.~Flambaum, J.~S.~M.~Ginges and M.~G.~Kozlov,
  ``Electric dipole moments of Hg, Xe, Rn, Ra, Pu, and TlF induced by the
  nuclear Schiff moment and limits on time-reversal violating  interactions,''
  Phys.\ Rev.\  A {\bf 66} (2002) 012111.
%  [arXiv:hep-ph/0203202].
  %%CITATION = PHRVA,A66,012111;%%

%\cite{Latha:2009nq}
\bibitem{Latha:2009nq}
  K.~V.~P.~Latha, D.~Angom, B.~P.~Das and D.~Mukherjee,
  ``Probing CP violation with the electric dipole moment of atomic mercury,''
  Phys.\ Rev.\ Lett.\  {\bf 103} (2009) 083001.
%  [arXiv:0902.4790 [physics.atom-ph]].
  %%CITATION = PRLTA,103,083001;%%

%\cite{Collaboration:2011tk}
\bibitem{Collaboration:2011tk}
  CMS~Collaboration,
  ``Search for Supersymmetry in pp Collisions at 7 TeV in Events with Jets and
  Missing Transverse Energy,''
  arXiv:1101.1628 [hep-ex].
  %%CITATION = ARXIV:1101.1628;%%

\bibitem{SPS}
M.~Battaglia {\it et al.},
``Proposed post-LEP benchmarks for supersymmetry,''
  Eur.\ Phys.\ J.\  C {\bf 22} (2001) 535;\\
%  [arXiv:hep-ph/0106204];
  %%CITATION = EPHJA,C22,535;%%
%\cite{Allanach:2002nj}
%\bibitem{Allanach:2002nj}
  B.~C.~Allanach {\it et al.},
``The Snowmass points and slopes: Benchmarks for SUSY searches,''
  arXiv:hep-ph/0202233;\\
  %%CITATION = ECONF,C01063;%%
%\cite{Ghodbane:2002kg}
%\bibitem{Ghodbane:2002kg}
  N.~Ghodbane and H.~U.~Martyn,
``Compilation of SUSY particle spectra from Snowmass 2001 benchmark
  models,''
%in {\it Proc. of the APS/DPF/DPB Summer Study on the Future of
%  Particle Physics
%(Snowmass 2001)
%} ed. N.~Graf,
  arXiv:hep-ph/0201233;\\
  %%CITATION = HEP-PH/0201233;%%
M.~Battaglia, A.~De Roeck, J.~R.~Ellis, F.~Gianotti, K.~A.~Olive and L.~Pape,
  ``Updated post-WMAP benchmarks for supersymmetry,''
  Eur.\ Phys.\ J.\  C {\bf 33} (2004) 273.
%  [arXiv:hep-ph/0306219].
  %%CITATION = EPHJA,C33,273;%%

%\cite{Regan:2002ta}
\bibitem{Regan:2002ta}
  B.~C.~Regan, E.~D.~Commins, C.~J.~Schmidt and D.~DeMille,
  ``New limit on the electron electric dipole moment,''
  Phys.\ Rev.\ Lett.\  {\bf 88} (2002) 071805.
  %%CITATION = PRLTA,88,071805;%%

%\cite{Baker:2006ts}
\bibitem{Baker:2006ts}
  C.~A.~Baker {\it et al.},
  ``An improved experimental limit on the electric dipole moment of the
  neutron,''
  Phys.\ Rev.\ Lett.\  {\bf 97} (2006) 131801.
  %[arXiv:hep-ex/0602020].
  %%CITATION = PRLTA,97,131801;%%

%\cite{Lebedev:2004va}
\bibitem{Lebedev:2004va}
  O.~Lebedev, K.~A.~Olive, M.~Pospelov and A.~Ritz,
  ``Probing CP violation with the deuteron electric dipole moment,''
  Phys.\ Rev.\  D {\bf 70} (2004) 016003.
  %[arXiv:hep-ph/0402023].
  %%CITATION = PHRVA,D70,016003;%%%\cite{Lebedev:2004va}




\end{thebibliography}
\end{document}